\documentclass[3p,twocolumn]{elsarticle}
\pdfoutput=1
\DeclareGraphicsExtensions{.pdf}

\biboptions{sort&compress}
\usepackage[colorlinks=true,linkcolor=blue, citecolor=blue, urlcolor=blue]{hyperref}
\usepackage{natbib}
\usepackage{color}
\usepackage{multirow}
\usepackage{amssymb}
\usepackage{amsmath}
\usepackage{slashed}
\usepackage{graphicx}
\usepackage{placeins}
\usepackage{array}
\usepackage{dblfloatfix}

\newcommand{\PbPb}{\mbox{Pb+Pb}}

\newcommand{\pp}{\mbox{$pp$}}

\newcommand{\Npart}{\mbox{$N_{\mathrm{part}}$}}

\newcommand{\fqref}{\mbox{${f_q}_0$}}
\newcommand{\fqint}{\mbox{$f_q^{\mathrm{int}}$}}

\newcommand{\PYTHIA}{PYTHIA8}

\newcommand{\RAA}{\mbox{$R_{\rm AA}$}}
\newcommand{\Raa}{\mbox{$R_{\rm AA}$}}

\newcommand{\qhat}{\mbox{$\hat{q}$}}

\newcommand{\GeV}{\mbox{${\rm GeV}$}}

\newcommand{\antikt}{\mbox{anti-\kt}}

\newcommand{\fqz}{\mbox{${f_q}_0$}}

\newcommand{\kt}{\mbox{$k_{t}$}}

\newcommand{\Dz}{\mbox{$D(z)$}}

\newcommand{\Dpt}{\mbox{$D(\pT)$}}

\newcommand{\pthat}{\mbox{$\hat{p}_{\mathrm{T}}$}}

\newcommand{\pt}{\mbox{$p_\mathrm{T}$}}
\newcommand{\pT}{\mbox{$p_\mathrm{T}$}}
\newcommand{\ptjet}{\mbox{$p_{{\mathrm{T}}}^{\mathrm{jet}}$}}
\newcommand{\pTjet}{\mbox{$p_{{\mathrm{T}}}^{\mathrm{jet}}$}}
\newcommand{\mathpTjet}{p_{\mathrm{T}}^{\mathrm{jet}}}
\newcommand{\pTz}{\mbox{${p_\mathrm{T}}_0$}}
\newcommand{\mathpTz}{{p_\mathrm{T}}_0}

\newcommand{\pTch}{\mbox{$p_{{\mathrm{T}}}^{\mathrm{ch}}$}}
\newcommand{\ptch}{\mbox{$p_{{\mathrm{T}}}^{\mathrm{ch}}$}}

\newcommand{\Rdz}{\mbox{$R_{D(z)}$}}
\newcommand{\Rdpt}{\mbox{$R_{D(\pT)}$}}

\newcommand{\fsoft} {\mbox{$\Phi^{\mathrm{soft}}$}}
\newcommand{\fsoftq}{\mbox{$\Phi_{q}^{\mathrm{soft}}$}}
\newcommand{\fsoftg}{\mbox{$\Phi_{\mathrm{g}}^{\mathrm{soft}}$}}
\newcommand{\fsoftinc}{\mbox{$\Phi_{\mathrm{inc}}^{\mathrm{soft}}$}}
\newcommand{\cf}            {C_\mathrm{F}}

\begin{document}

\begin{frontmatter}
\title{
\center{Interpreting Single Jet Measurements in \PbPb\ Collisions at the LHC}
}

\author[Charles]{Martin Spousta}
\author[Columbia]{Brian Cole}
\address[Charles]{Institute of Particle and Nuclear Physics, Charles University, Prague, Czech Republic}
\address[Columbia]{Columbia University Physics Department and Nevis Laboratories, New York, NY, 10027 USA}

\begin{abstract}
Results are presented from a phenomenological analysis of recent
measurements of jet suppression and modifications of jet fragmentation
functions in \PbPb\ collisions at the LHC. Particular emphasis is
placed on the impact of the differences between quark and gluon jet
quenching on the transverse momentum (\pTjet) dependence of the jet \RAA\ and
on the fragmentation functions, $D(z)$. Primordial quark and gluon
parton distributions were obtained from PYTHIA8 and were parameterized
using simple power-law functions and extensions to the power-law
function which were found to better describe the PYTHIA8 parton
spectra. A~simple model for the quark energy loss based on the shift
formalism is used to model \RAA\ and \Dz\ using both analytic results
and using direct Monte-Carlo sampling of the PYTHIA parton
spectra. 
  The model is capable of describing the full \pTjet , rapidity, and 
centrality dependence of the measured jet \RAA\ using three effective 
parameters.
  A key result from the analysis is that the 
\Dz\ modifications observed in the data, excluding the enhancement at
low-$z$, may result primarily from the different quenching of the
quarks and gluons. The model is also capable of reproducing the
charged hadron \RAA\ at high transverse momentum. Predictions are made
for the jet \RAA\ at large rapidities where it has not yet been
measured and for the rapidity dependence of \Dz.

\end{abstract}
\end{frontmatter}

\newlength{\fighalfwidth}
\setlength{\fighalfwidth}{0.49\textwidth}

\section{Introduction}
Measurements of jet production and jet properties in
ultra-relativistic nuclear collisions provide an important
tool to study the properties of quark gluon plasma created in the
collisions. High-energy quarks and gluons produced in
hard-scattering processes can interact with and lose energy while
propagating in the plasma. Those interactions can both reduce the
energy of the jets that result from the fragmentation of the quarks
and gluons and change the properties of the jets. These and other
``medium'' modifications of the parton showers initiated by the hard
scattering \cite{Mehtar-Tani:2013pia,Armesto:2011ht} are frequently
collectively referred to as ``jet quenching''. 
\begin{table*}[!b]
\begin{center}
\begin{tabular}{|l|l||c|c|c|c|} \hline
Fit type            & Parameter  & $|y|<2.1$ & $|y| < 0.3$ & $0.3<|y|<0.8$ 	& $1.2<|y|<2.1$	\\ \hline  \hline
All                 & \fqz\      & 0.34      & 0.28   & 0.29 & 0.40
\\ \hline   \hline
\multirow{2}{*}{Power law} & $n_q$    & 5.66      & 5.37	   & 5.40		& 6.15		\\ \cline{2-6}
                  & $n_{g}$  & 6.25      & 5.97	   & 6.09     & 6.92
\\ \hline \hline
                  & $n_q$    & 4.19 & 4.34 & 4.27 & 3.75     \\ \cline{2-6}
Extended          &  $\beta_{q}$          & 0.71 & 0.49 & 0.54 & 1.2   \\ \cline{2-6}
power law         &  $n_{g}$              & 4.69  & 4.55 & 4.57 & 4.60     \\ \cline{2-6}
                  &  $\beta_{g}$          & 0.80 & 0.71 & 0.76 & 1.2   \\ \hline
\end{tabular}
\end{center}
\caption{Parameters obtained from fits of the \PYTHIA\ jet spectra to power-law (Eq.~\ref{eq:powerlaw}) 
and extended power-law (Eq.~\ref{eq:extpowerlaw}) functions.
}
\label{tab:fitparams}
\end{table*}

Jet quenching was first observed at the LHC through the observation of
highly asymmetric dijet pairs \cite{Aad:2010bu} that result when the two
jets lose different amounts of energy in plasma. Since dijet pairs for
which both jets lose similar energy or, more generally, have similar 
modifications will appear ``symmetric'', other observables are needed
to probe the effects of quenching on the typical jet. Measurements of
the suppression of the hadron spectrum resulting from the energy loss
of the parent jets have been carried out at both RHIC
\cite{Adams:2003kv,Adler:2003au,Adare:2008qa} and the 
LHC \cite{Abelev:2012hxa,CMS:2012aa,ATLAS:2012dna}. These show a
suppression that at the LHC varies from a 
factor $\sim 5$ for hadron transverse momentum (\pT) values  $\sim
10$~GeV to a factor of $\sim 2$ for $\pT \gtrsim 50$~\GeV. Most
jet quenching calculations that attempt to infer medium properties
such as the quenching transport parameter, \qhat, (see e.g. \cite{Burke:2013yra}
and references therein) have relied on the single hadron suppression
results because of the theoretical simplicity in calculating single
hadron spectra. However, the single hadron measurements have only
indirect sensitivity to the kinematics of the parent parton and little
sensitivity to the details of the modification of the parton shower.

Recent measurements of the suppression of the jet yield in
2.76~TeV \PbPb\ collisions at the LHC \cite{Aad:2014bxa} are expected to
provide a more sensitive probe of the physics of jet quenching at
least through the improved correlation between the measured jet and the
parent parton (shower) kinematics. Recent measurements of the jet
nuclear modification factor, \RAA, for high transverse momentum jets
show a factor of $\sim 2$ suppression in the jet yield that increases 
slowly with increasing \pTjet. The suppression is observed to vary
monotonically as a function of collision centrality and to be
independent of jet rapidity within the statistical and systematic
uncertainties. Separately, recent measurements of
the fragmentation functions for jets produced in \PbPb\ collisions
\cite{Aad:2014wha,Chatrchyan:2014ava} have shown an enhanced yield of hadrons with low transverse
momenta, \pT, or low jet momentum fraction, $z$, a suppressed yield of
hadrons with $0.05 \lesssim z \lesssim 0.2$ and possibly an enhanced
yield of hadrons with $z \gtrsim 0.2$. These two observables, the
single jet suppression and the modification of the fragmentation
function, arguably provide the minimal set of data needed to
understand the physics of jet quenching: the jet suppression is
sensitive to the amount of energy the jet loses -- outside the jet
``cone'' -- while the fragmentation function is sensitive to the
re-distribution of energy inside the jet cone. 

Interpretation of the inclusive jet suppression and fragmentation data
is complicated by the flavor admixture of the primordial partons.  For
this analysis, we focus on the relative combination of light quarks
and gluons which are expected to suffer different energy loss due to
their different color charges and/or differences between the quark
and gluon splitting functions. In weak coupling calculations, the
relative quark and gluon energy loss rates are determined by
perturbative QCD color factors. For example, it is usually assumed
that gluons lose energy at a rate $9/4$ higher than that for
quarks. Recent studies including those based on quenching Monte Carlo
codes have been compared to the jet \RAA\ and fragmentation function
measurements, but these analyses have not explicitly attempted to
elucidate the role of the relative quark and gluon contributions to
the jet spectrum.

In the remainder of this paper, we attempt to interpret recent
measurements of the single jet suppression and fragmentation function
ratios explicitly accounting for the role of the quark and gluon
admixture. We argue that some of the features of the data can be
explained purely on the basis of the \pTjet\ dependence of the quark
to gluon fraction of the primordial parton spectrum and the different
quark and gluon energy loss. Our analysis is based on simple
assumptions regarding the parametric dependence of the energy loss on
the jet transverse momentum and flavor. These assumptions are
sufficiently simple that the results of our analysis can be easily
explained and understood, but their simplicity also means that the
results presented here should be verified using a proper jet quenching
calculation. Our analysis is based on a combination of analytic
calculation and Monte Carlo simulation using simulated quark and gluon
spectra obtained from PYTHIA8 \cite{Sjostrand:2014zea}.

\section{Parameterizing Jet Spectra and \Dz\ distributions}
\label{sec:JetSpectra}
\begin{figure*}[!t]
\centerline{
\hspace*{\fill}
\includegraphics[width=0.9\fighalfwidth]{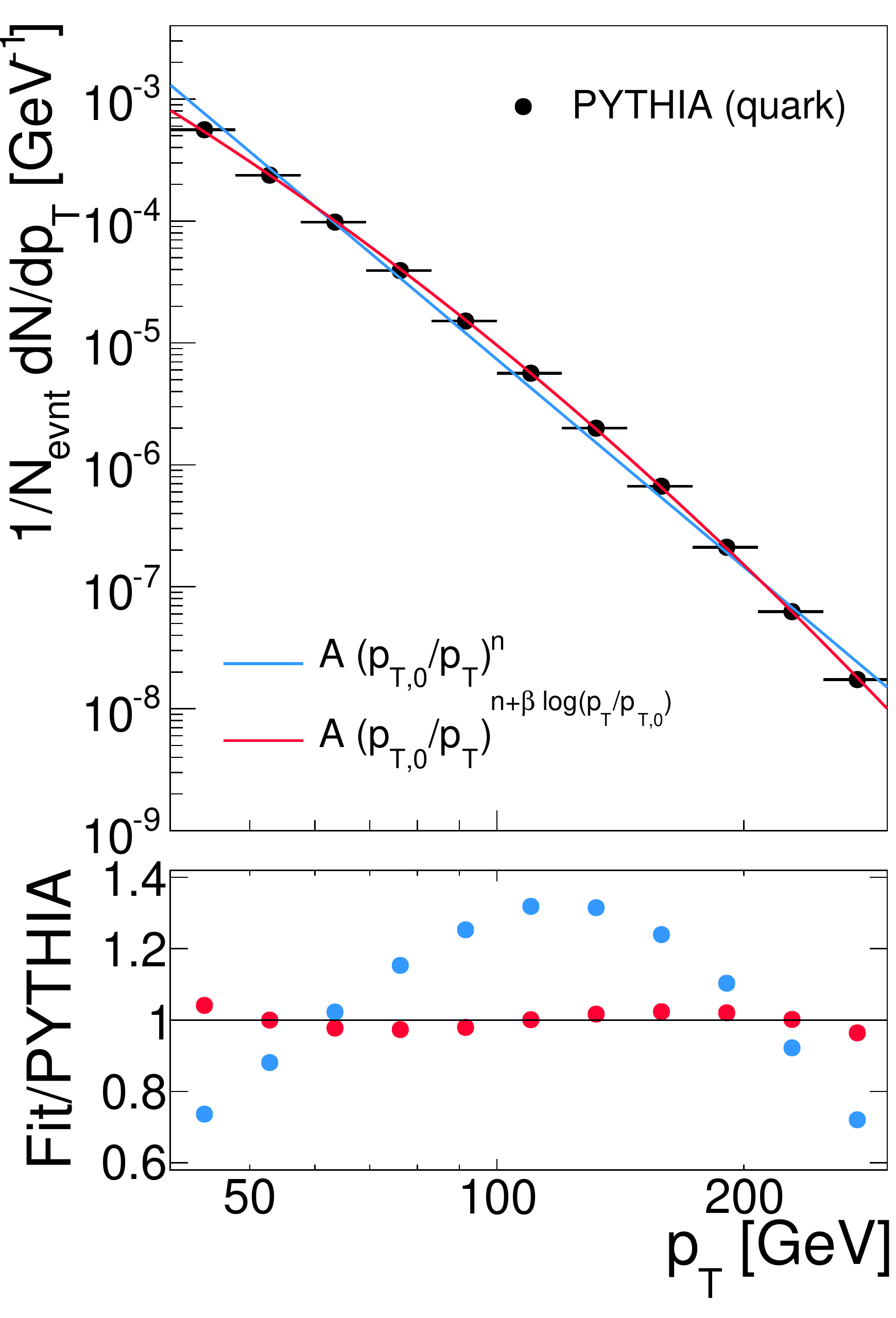}
\hfill
\includegraphics[width=0.9\fighalfwidth]{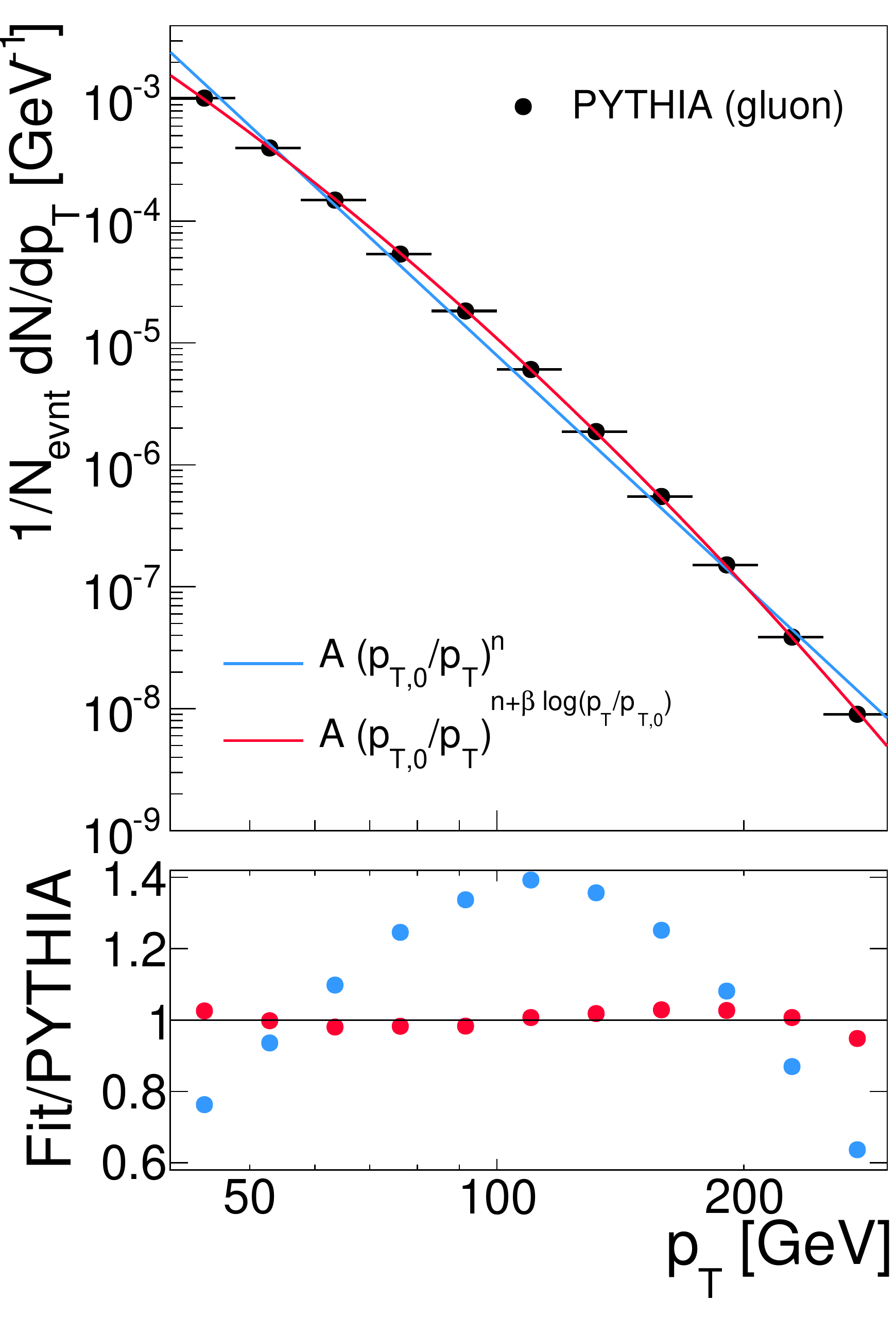}
\hspace*{\fill}
}
\caption{
Top panels: Quark (left) and gluon (right) spectra obtained from
\PYTHIA\ simulations over the rapidity interval $|y| < 2.1$ with
simple power law (Eq.~\ref{eq:powerlaw}) and modified power law
(Eq.~\ref{eq:extpowerlaw}) fits superimposed. Bottom panels: ratios
of the spectra to the two fit functions. 
}
\label{fig:inclspectra}
\end{figure*}

A key ingredient of the analysis in this paper is the quark and gluon 
jet spectra which were obtained from \PYTHIA\ using procedures chosen to 
be similar to those used in certain ATLAS simulations 
\cite{Aad:2012vca,Aad:2014hia}. Namely, \PYTHIA\ was run using 
parameters from the AU2 tune \cite{ATLAS:2012uec} and using CT10 parton 
distribution functions \cite{Lai:2010vv}. This combination was shown to 
describe well the LHC jet data \cite{Aad:2014hia}.
   Jets were reconstructed using the
\antikt\ algorithm \cite{Cacciari:2008gp} with the distance
parameter $R=0.4$ applied to hadrons with lifetimes $c\tau >
1$~mm. The resulting jets were matched to one of the two
outgoing partons from the leading order hard-scattering process by
choosing the parton with the smallest angular distance, $\Delta R =
\sqrt{\Delta \phi^2 + \Delta \eta^2}$ and the flavor of the jet was
assigned to be that of the matched parton.  
  Six million PYTHIA hard-scattering events were generated for each of five 
intervals of $\pthat$, the transverse momentum of outgoing
partons in the $2 \rightarrow 2$
hard-scattering, with boundaries 17, 35, 70, 140, 280 and 560~GeV.

The Monte Carlo results presented in this paper were obtained by directly
using the jets obtained from the \PYTHIA\ simulations, but the analytic
results require a parameterization of the \pTjet\ dependence of the
jet yields. A common parameterization used to describe the spectra of
high-\pT\ hadrons or jets is the power-law form, e.g.
\begin{equation}
\frac{dn}{d\pTjet} = A \left(\frac{\pTz}{\pTjet}\right)^n,
\label{eq:powerlaw}
\end{equation}
where \pTz\ is a reference transverse momentum value at which $A$
represents $dn/d\pTjet$. Results of fits to the quark and gluons
distributions from the $|y|<2.1$ rapidity interval are shown in
Fig.~\ref{fig:inclspectra}. The values of $n$ extracted from the pure
power-law fits are listed in Table~\ref{tab:fitparams}. The power-law
function can describe the gross-features of the spectra but the ratios of the spectra to the
fit functions presented in the bottom of the figure indicate
significant deviations of the jet spectra from the power-law form. 

The power-law distribution can be improved by adding a logarithmic
\pTjet\ dependence to the exponent producing an ``extended power-law'', 
\begin{equation}
\frac{dn}{d\pTjet} = A \left(\frac{\pTz}{\pTjet}\right)^{n + \beta \log{\left( \mathpTjet/\mathpTz \right)}}.
\label{eq:extpowerlaw}
\end{equation}
With this form, $\beta$ represents the logarithmic derivative of
$dn/d\mathpTjet$ at $\pTjet = \pTz$. At the most forward rapidities,
the strong phase-space suppression of the jet spectra at high
\pTjet\ makes even the extended power-law inadequate for describing
the jet spectra. Thus, for the most forward rapidities, an additional
quadratic term, $\gamma \log^2{\left( \ptjet/ \pTz \right)}$, is added to
the power-law exponent and the resulting function is capable of
describing the most forward quark and gluon spectra over the
\pTjet\ range used in this analysis.

\begin{figure}[t]
\centerline{
\includegraphics[width=0.8\fighalfwidth]{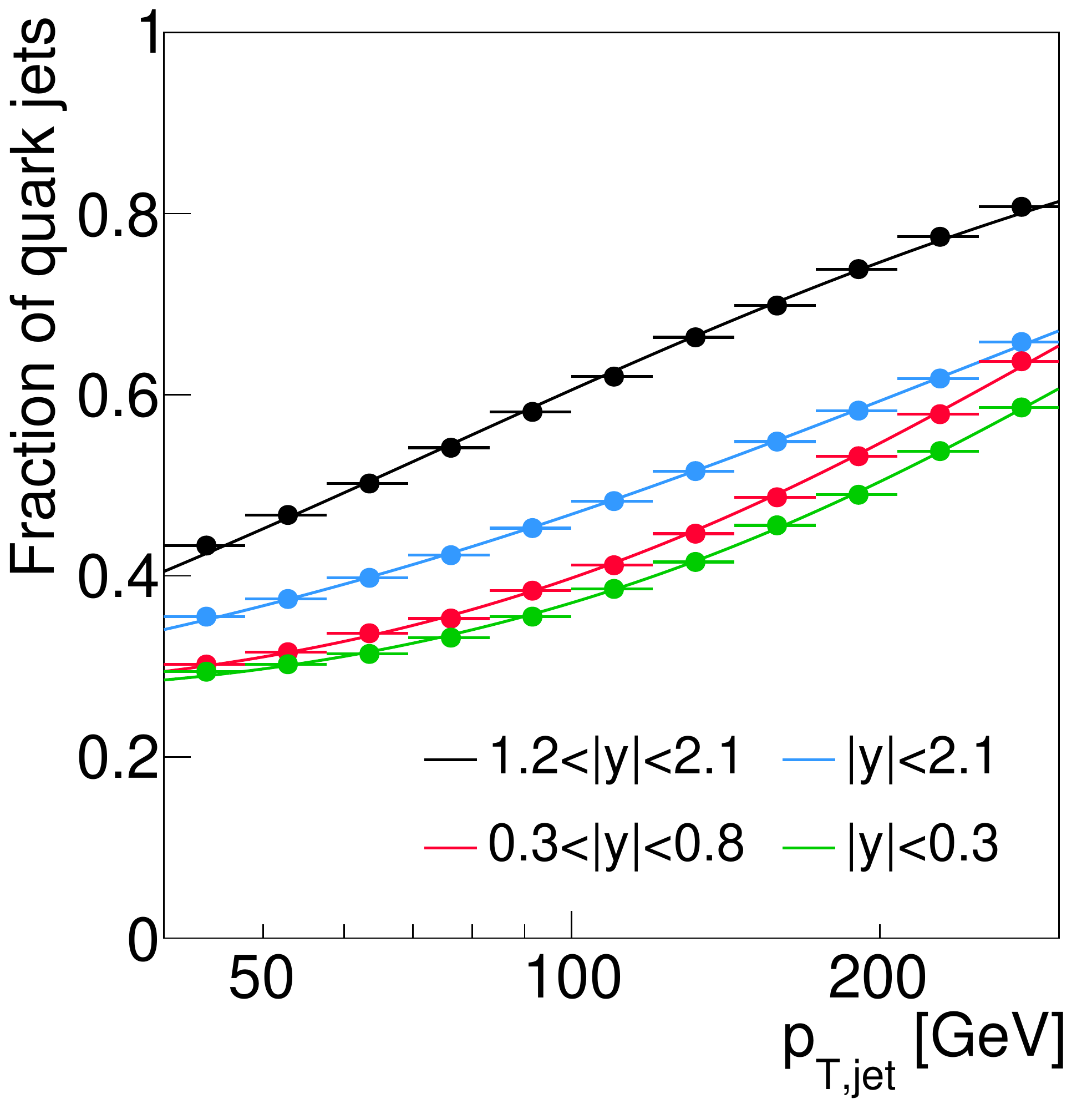}
}
\caption{
Jet quark fraction as a function of \pTjet\ in the different jet
rapidity intervals used in this study. The points show results
obtained from \PYTHIA\ simulations, the solid lines represent results
obtained from extended power-law fits with the parameters shown in
Table~\ref{tab:fitparams}. 
}
\label{fig:fqpt}
\end{figure}

A jet spectrum that consists of a mixture of quark and gluon
contributions can be represented in terms of a sum of contributions
each of the form of Eq.~\ref{eq:powerlaw} or its extensions. However,
for the purposes of this paper, it will be convenient to express the
combined spectrum in terms of a quark fraction, \fqz, specified at
\pTz. Then a combined spectrum using power-law forms can be written
\begin{equation}
 \frac{dN}{d\pTjet} =  A \left[ \fqz \left( \frac{\pTz}{\pTjet}\right)^{n_q}
  + \left(1 - \fqz\right)\left( \frac{\pTz}{\pTjet}\right)^{n_g}\right],
\label{eq:unmod}
\end{equation}
where $n_q$ and $n_g$ are the quark and gluon power-law indices,
respectively. Since $n_q \neq n_g$, the quark fraction will evolve as
a function of \pTjet\ according to 
\begin{equation}
\begin{split}
f_{q}\left(\pTjet\right) & = \frac{\fqz \left(\frac{\pTz}{\pTjet}\right)^{n_q}}{\fqz
  \left(\frac{\pTz}{\pTjet}\right)^{n_q} + \left(1 -
  \fqz\right)\left(\frac{\pTz}{\pTjet}\right)^{n_g}}\\
& = \frac{1}{1 + \left(\frac{1 - \fqz}{\fqz}\right)\left(
 \frac{\pTz}{\pTjet}\right)^{n_g - n_q}}.
\end{split}
\label{eq:fqpt}
\end{equation}
For the extended power-law parameterizations of the spectra, the \pTjet-dependent 
quark fraction looks similar to that in Eq.~\ref{eq:fqpt} but with the addition of a 
term, $\left(\beta_g - \beta_q\right) \log{\left(\pTjet/\pTz\right)}$ to the 
exponent in the denominator. The \pTjet\ dependence of the quark fraction is shown in Fig.~\ref{fig:fqpt}.

\begin{figure}[t]
\centerline{
\includegraphics[width=0.8\fighalfwidth]{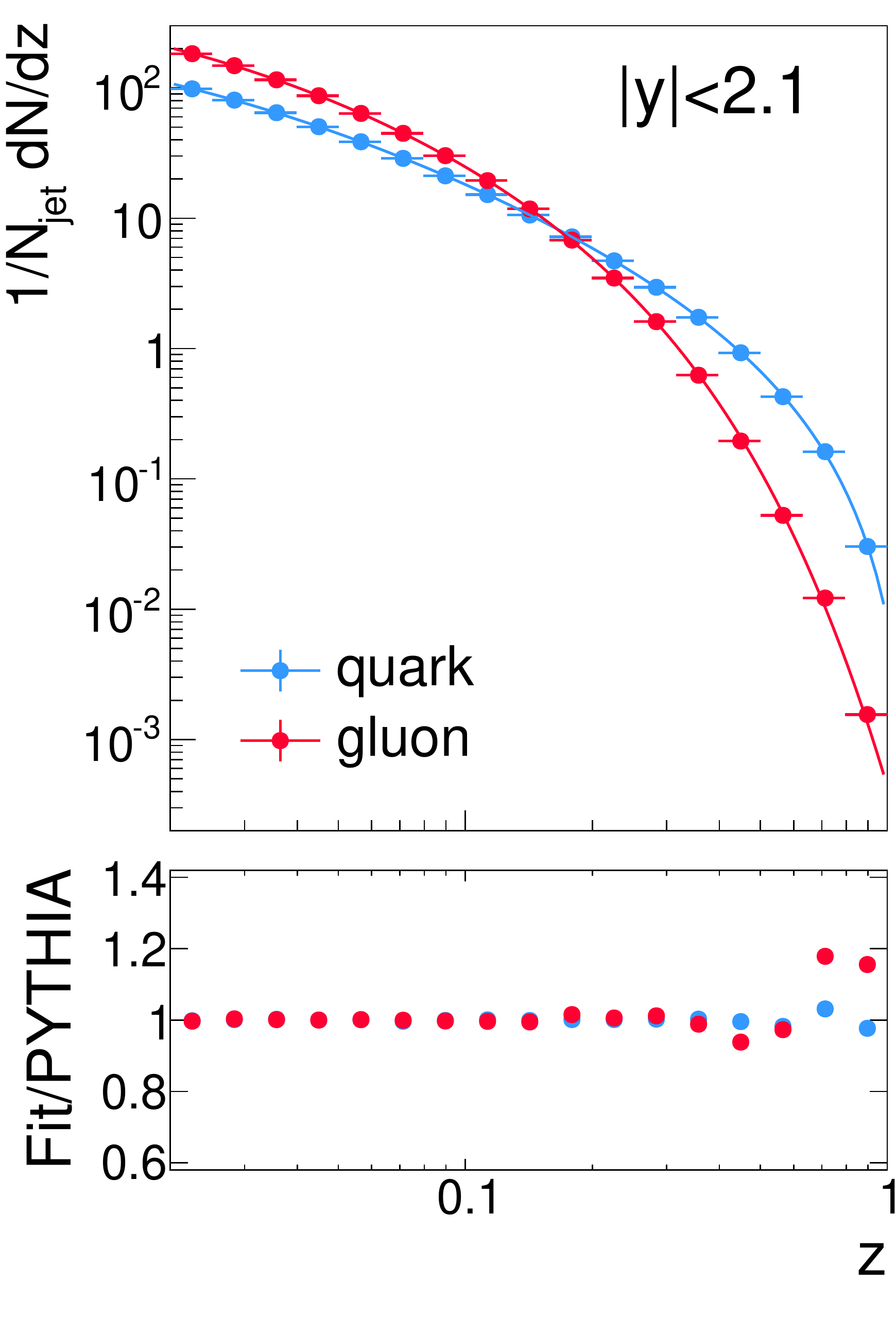}
}
\caption{
\PYTHIA\ quark and gluon \Dz\ distributions for $R = 0.4$ jets having
$\pTjet > 100$~GeV and $|y| < 2.1$. The solid lines show the
results of fits to the \Dz\ distributions using the function in Eq.~\ref{eq:dzfit}.
}
\label{fig:incldz}
\end{figure}

The \PYTHIA\ \Dz\ distributions were obtained using
final-state charged hadrons located within an angular radius, $\Delta
R < 0.4$, of reconstructed jets having $\pTjet > 100$~GeV. The
resulting distributions are shown in Fig.~\ref{fig:incldz} for the
rapidity interval $|y| < 2.1$. The quark
\Dz\ distribution is noticeably harder than the gluon
\Dz\ distribution, but is also lower at intermediate $z$, in the range
where the \Dz\ distribution appears to be depleted in \PbPb\ collisions. 

For use in the analytic analysis, the \Dz\ distributions were fit to
functions of the form, 
\begin{equation}
\Dz = a \cdot \frac{ (1+ d z)^b } { (1 + e z)^c } \cdot \exp{(- f z)}
\label{eq:dzfit}
\end{equation}
which are similar to other commonly used parameterizations \cite{deFlorian:2007aj}
with the addition of an exponential term. That term is not used for
the quark distributions, but it's presence provides a more controlled
description of the gluon \Dz\ distribution. 
The results of the fits for the quark and gluon distributions 
over $|y|<2.1$ are shown in Fig.~\ref{fig:incldz}, and the ratios of
the fit to the \PYTHIA\ \Dz\ distributions are shown in the lower
panels. The fits well describe the simulated \Dz\ distributions with
parameters that are provided in Table~\ref{tab:dzfitparams}. We note
that the parameterization in Eq.~\ref{eq:dzfit} has a smooth
extrapolation past $z = 1$. The pQCD fragmentation function has no
contribution from $z > 1$, but when reconstructing jets in
\PYTHIA\ and data, there are events having two jets that are close
enough that a high-\pTch\ fragment from the higher-energy jet can be
associated with the lower-energy jet possibly yielding a hadron with
$z > 1$. The \Dz\ distributions fall rapidly above $z = 1$ so they have no
practical importance, though the continuity of the parameterization
will be relevant later in this paper.

\begin{table}
\begin{center}
\begin{tabular}{|l||c|c|c|c|c|c|} \hline
    & $a$ & $b$ & $c$ & $d$ & $e$ & $f$  \\ \hline
quark & 318 & 2.51 & 1.44 & -0.85 & 52.4 & 0  \\ \hline
gluon & 574 & 1.87 & 2.32 &  9.09 & 32.0 & 10.3 \\ \hline
\end{tabular}
\end{center}
\caption{
Parameters describing the fragmentation functions extracted from
\PYTHIA\ using the procedures described in the text for the functional
form in Eq.~\ref{eq:dzfit}.
}
\label{tab:dzfitparams}
\end{table}

\section{Analytic models for jet \RAA\ and \Dz: power-law spectra}
\label{sec:RAADZConstantfracShift}
Our modeling of the single jet suppression is based on the ``shift''
approach of \cite{Baier:2001yt} in which the quenched jet spectrum can be
(approximately) written
\begin{equation}
\dfrac{dn_{\mathrm{Q}} (\pTjet)}{d\pTjet}  = \dfrac{dn \left(
\pTjet + S(\pTjet) \right)}{d\pTjet}   \times \left(1 + \frac{dS}{d\pTjet}\right),
\label{eq:quenchspect}
\end{equation} 
where $dn_{\mathrm{Q}}$ and $dn$ represent the per-event yields of quenched
and unquenched jets. 
The second term in Eq.~\ref{eq:quenchspect} is a Jacobian term that is
necessary to (e.g.)  preserve the total number of jets.

Using the power-law form for a single-flavor jet spectrum, 
\begin{equation}
\dfrac{dn_Q (\pTjet)}{d\pTjet}  = A \left(\dfrac{\pTz}{\pTjet + S(\pTjet)}\right)^n  \left(1 + \frac{dS}{d\pTjet}\right).
\label{eq:spectmodpowerlaw}
\end{equation} 
The ratio of the quenched and unquenched spectra, the analog of the
measured \RAA, is then
\begin{equation}
\RAA (\pTjet)  =  \left(\frac{1}{1 + S(\pTjet)/\pTjet}\right)^n   \left(1 + \frac{dS}{d\pTjet}\right).
\end{equation} 
It has been previously observed \cite{Adcox:2004mh} \footnote{The
  definition of $n$ differs between this paper and \cite{Adcox:2004mh}
  where it characterizes the invariant cross-section and is,
  therefore, larger by one due to the $dp_{\mathrm{T}}^2$ factor. 
} 
that if the shift is proportional to
\pTjet, $S \equiv s \pT$, then the resulting \RAA\ is \pTjet-independent:
\begin{equation}
\RAA (\pTjet) = \frac{1}{\left (1 + s\right)^{n-1}},
\end{equation} 
such that the fractional shift can be inferred from an approximately
\pTjet-independent \RAA\ \cite{Adcox:2004mh}
\begin{equation}
s = \frac{1}{\RAA^{\left(\frac{1}{n - 1}\right)}} - 1. 
\label{eq:simpleshift}
\end{equation} 
This result has previously been applied to single hadron
\RAA\ measurements but is arguably more appropriate when applied to
jet \RAA\ measurements. Then, naively applying
Eq.~\ref{eq:simpleshift} to the typical suppression observed at high
\pTjet\ in central \PbPb\ collisions at the LHC, $\RAA \sim 0.5$, and
using $n \sim 5$, the resulting fractional shift would be $s =
0.19$. The observation that the jet \RAA\ is only weakly dependent on
\pTjet\ at high \pTjet\ has been taken as evidence that jets lose a
constant fraction of their energy in the quark gluon plasma created in
\PbPb\ collisions. There are potentially significant theoretical flaws
with this conclusion, but the conclusion and the interpretation of
the extracted $s$ also suffer from neglecting the fact that the jet
spectrum is composed of an admixture of flavors.

Starting with a combination of quark and gluon power-law spectra
(Eq.~\ref{eq:unmod}), the quenched spectrum would be
\begin{equation}
\begin{split}
 \frac{dN_{\mathrm{Q}}}{d\pTjet} & =  A \left[ \fqref \left( \frac{\pTz}{\pTjet + S_q}\right)^{n_q}
 \left(1 + \frac{dS_q}{d\pTjet}\right) + \right.\\
 & \phantom{{}={}} \left.\left(1 - \fqref\right)\left(
 \frac{\pTz}{\pTjet + S_g}\right)^{n_g}\left(1 + \frac{dS_g}{d\pTjet}\right)\right].
\end{split}
\label{eq:modpower}
\end{equation}
The resulting \RAA\ would be given by the ratio of Eq.~\ref{eq:modpower} to
Eq.~\ref{eq:unmod} which, with some simplification, takes the form
\begin{equation}
\begin{split}
 \RAA & = f_q \left(\frac{1}{1 + S_q/\pTjet}\right)^{n_q}   \left(1 + \frac{dS_q}{d\pTjet}\right) +\\
& \phantom{{}={}}  \left(1 - f_q\right) \left(\frac{1}{1 + S_g/\pTjet}\right)^{n_g}   \left(1 + \frac{dS_g}{d\pTjet}\right).
\end{split}
\label{eq:raaqg}
\end{equation}
Here, $f_{\mathrm{q}}$ is the full \pTjet-dependent quark fraction in Eq.~\ref{eq:fqpt}.
As the equation indicates, the combined jet \RAA\ is given by a
combination of the separate quark and gluon suppression factors
weighted by the quark and gluon fractions. 
\begin{figure*}[t]
\centerline{
\includegraphics[width=0.8\textwidth]{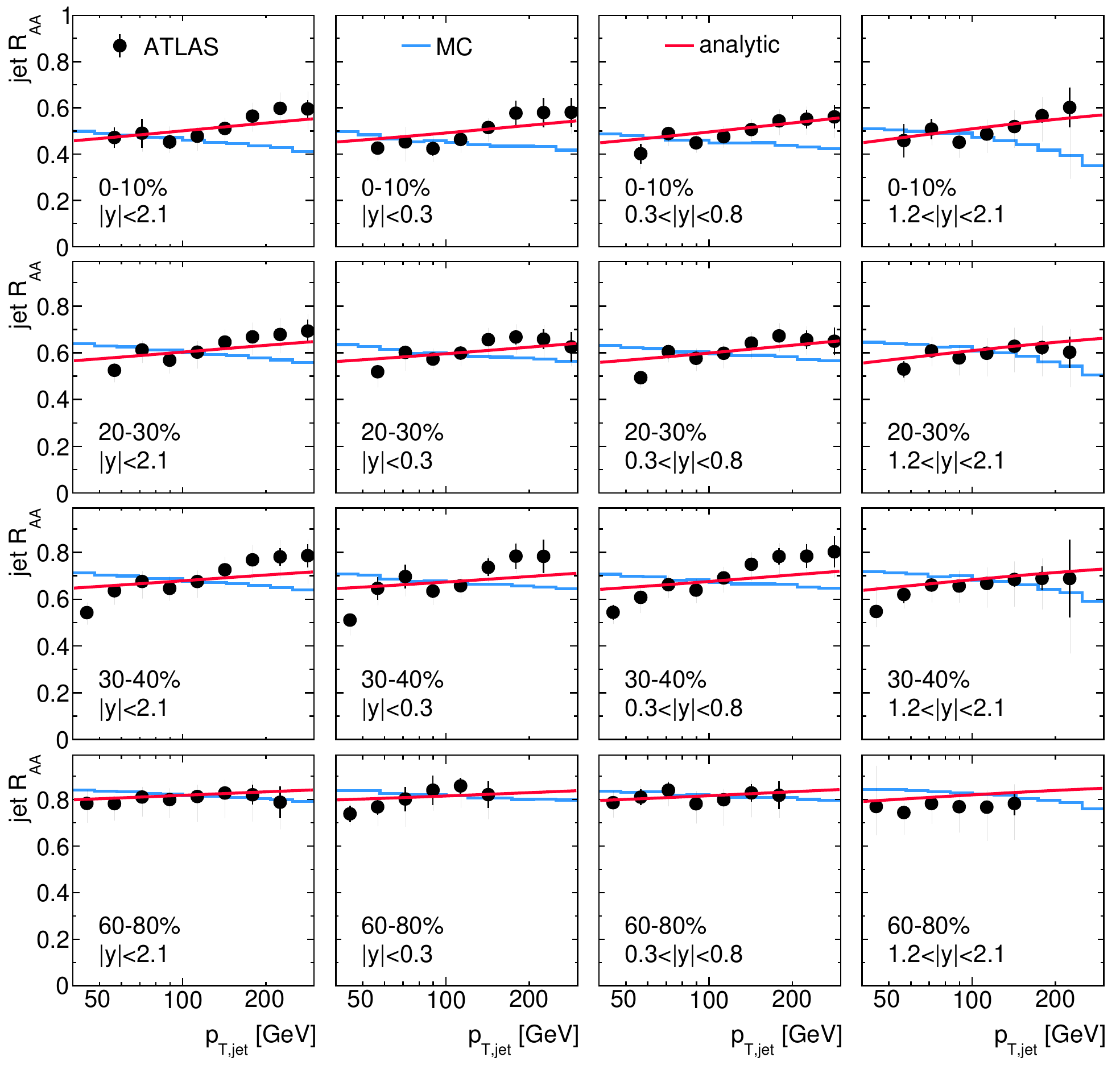}
}
\caption{
Nuclear modification factor of jets, \Raa, measured by ATLAS \cite{Aad:2014bxa}
(black markers) in four different centrality bins (rows) and four
different rapidity regions (columns) compared to the analytic
calculation (red line) and MC calculation (blue histogram) of 
the \RAA\ using constant fractional shift and power-law spectra.
}
\label{fig:raacomparepower}
\end{figure*}

Even in the case of
constant fractional energy loss for the quarks and gluons, $S_q = s_q
\pTjet$ and $S_g = s_g \pTjet$, Eq.~\ref{eq:raaqg} would imply an
\RAA\ that varies with \pTjet\ as long as $S_q \neq S_g$. For example,
if $S_g/S_q > 1$, then the \RAA\ will increase with \pTjet\ because of
an increasing quark fraction and weaker suppression for the
quarks. Such an increase of \RAA\ with increasing \pTjet\ has been
observed in the measured \RAA\ values 
that are shown in Fig.~\ref{fig:raacomparepower}. To test whether the
data are compatible with the constant fractional energy loss scenario,
we have assumed  $s_g = 9/4 \times s_q$ and have fit the \RAA\ values using
Eq.~\ref{eq:raaqg} with one free parameter for each
centrality bin, namely $s_q$. The results are
shown in Fig.~\ref{fig:raacomparepower} with solid lines. The
extracted $s_q$ values vary from 0.02 for the 60-80\% centrality bin
to 0.1 for the 0-1\% bin. The figure shows that the constant
fractional shift assumption combined with the power-law form for the
jet spectra is capable of approximately reproducing the slow variation
of the measured jet \RAA\ with \pTjet. 

Since the \pTjet\ dependence of the jet suppression can be
successfully explained using a combination of the varying quark
fraction and the greater quenching of gluon jets, it is worth
exploring the impact of these same behaviors on the jet fragmentation
function. In particular, we wish to determine whether the different
quenching of the quarks and gluons can explain part or all of the the
observed modifications of the fragmentation function in
\PbPb\ collisions. We make the simplest possible assumption, namely
that the quarks and gluons lose energy in the plasma and then fragment
according to vacuum fragmentation functions. With this assumption, the
only source of modification to the inclusive jet fragmentation
function is the change in the quark (or gluon) fraction due to the
medium-induced energy loss. Starting from Eq.~\ref{eq:modpower}, the
modified quark fraction in the constant fractional shift scenario is
\begin{equation}
f_q^{\mathrm{mod}} = \frac{1}{1 + \left(\frac{1 - \fqz}{\fqz}\right)
\dfrac{\left(1 + s_g\right)^{n_g - 1}}{\left(1 + s_q\right)^{n_q - 1}}
\left(\frac{\pTz}{\pTjet}\right)^{n_g - n_q}}.
\end{equation}

Assuming that the \Dz\ distributions are independent of \pTjet, the
per-jet distribution of fragments as a function of the fragment
longitudinal momentum fraction, $D(z)$, can be written, 
\begin{equation}
D(z) = f_q^{\mathrm{int}} D_q(z) + (1 - f_q^{\mathrm{int}}) D_g(z),
\label{eq:dzmodpower}
\end{equation}
where $D_q(z)$ and $D_g(z)$ are the quark and gluon $D(z)$
distributions, respectively, and $f_q^{\mathrm{int}}$ is the modified
quark fraction integrated over a given \pTjet\ range. 

\begin{figure*}[t]
\centerline{
\includegraphics[width=0.9\textwidth]{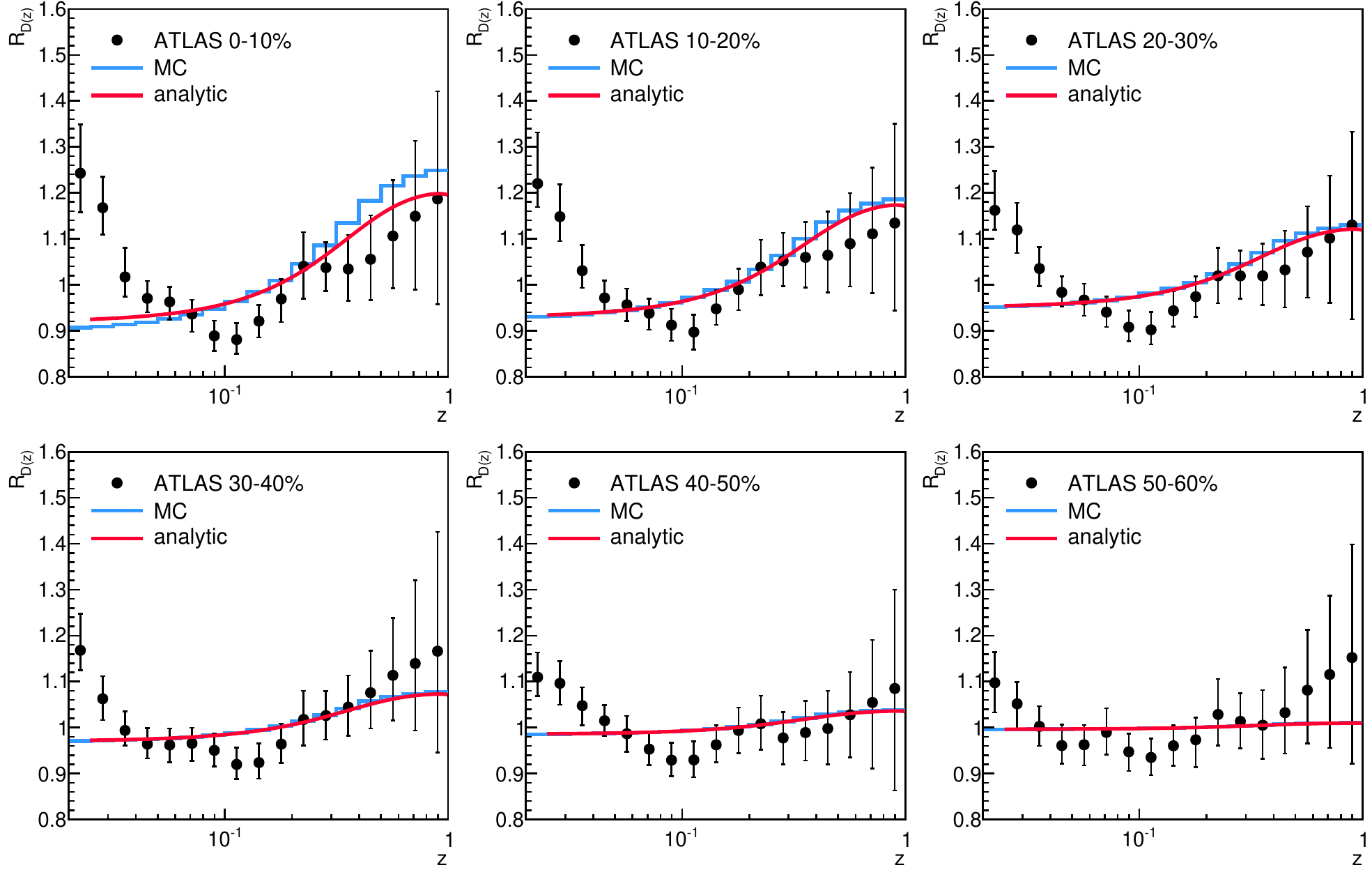}
}
\caption{
Ratios of $D(z)$ distributions for six bins in collision centrality to
those in peripheral (60-80\%) collisions, $D(z)|_{\mathrm{cent}}
/D(z)|_{60-80}$, measured by ATLAS for $R=0.4$ jets \cite{Aad:2014wha} (black
markers) are compared to the analytic calculation (red line) and MC
calculation (blue histogram) of the same quantity in the fractional
energy loss model. The analytic calculation uses the power law
parameterization of jet \pt\ spectra.
}
\label{fig:dzcomparepower}
\end{figure*}

The ATLAS jet fragmentation measurements were obtained for $\pTjet >
100$~GeV. Applying Eq.~\ref{eq:dzmodpower} over this \pTjet\ range
and using the $s_q$ parameters obtained from fits to the jet \RAA, we
calculated the ratio of modified \Dz\ distributions in different
centrality bins to the distribution in the 60-80\% centrality bin
for comparison with the ATLAS data. The results are shown along with
the data in Fig.~\ref{fig:dzcomparepower}. The figure shows that our
simple model for the medium modifications of the inclusive jet
fragmentation function can reproduce some of the qualitative features
in the data, namely the suppression of the fragmentation function at
intermediate $z$ and an enhancement in the fragmentation function at
large $z$. This latter is statistically marginal in the data given the
(combined) error bars, but the enhancement at large $z$ in the model
is an automatic result of the increased quark content of the jet
spectrum. Our model does not show as deep a suppression in the
\Dz\ ratio near 0.1 which may indicate that additional physics
contributes there. 

One feature in the data that cannot be explained by the
model is the enhancement at low $z$. Our simple model also explains the
centrality dependence of the data, except for the 50-60\%
centrality bin, given the fits to the single-jet suppression. 
Based on the results shown in Fig.~\ref{fig:dzcomparepower} we argue
that it is plausible that the modifications observed at intermediate and large
$z$ in the jet fragmentation function result from quenching-driven
changes in the jet quark fraction while the enhancement at low $z$ reflects
a contribution of extra particles in the jet either from radiative emission
within the jet or recoil of particles in the medium. 

We have performed a separate Monte-Carlo evaluation of the single-jet
suppression to check and improve on the results of the above analytic
calculations which are necessarily limited by assumptions regarding
the shapes of the jet spectra. To simulate the single-jet
suppression, we sample jets from the \PYTHIA-simulated events, apply 
the shift as in Eq.~\ref{eq:quenchspect} with chosen $S_q$ and $S_g$ for quark and gluon
jets, respectively, and then build the resulting spectra of quenched jets.
The simulated \RAA\ is obtained from the ratio of the quenched
spectrum to the original spectrum of \PYTHIA\ jets. The results are
shown with the blue histograms in Fig.~\ref{fig:raacomparepower}. The
agreement with the analytic results is poor, suggesting that the
power-law parameterization of the jet spectra is inadequate for the
simulation of the single-jet suppression. In fact, the Monte-Carlo
sampled \RAA\ decreases with increasing \pTjet\ in contradiction with
the general result that constant fractional
shift combined with the increasing quark fraction should produce an
\RAA\ that increases with \pTjet. The decrease of the Monte-Carlo
sampled \RAA\ must necessarily result from the jet spectra being
steeper, or having greater curvature than the power-law function or,
equivalently, from the differences between the power-law fit
functions and the \PYTHIA\ spectra seen in Fig.~\ref{fig:inclspectra}.

\section{Analytic models for jet \RAA: extended power-law spectra}
To test this conclusion, we have extended the analytic analysis from
the previous section to the case of extended power-law spectra. The
analog of Eq.~\ref{eq:spectmodpowerlaw} is  
\begin{equation}
\frac{dn_{\mathrm{Q}}}{d\mathpTjet}  = A \left(\dfrac{\mathpTz}{\mathpTjet +S}\right)^{n+\beta\log{\left[(\mathpTjet + S)/\mathpTz \right]}}  \left(1 + \frac{dS}{d\mathpTjet}\right),
\label{eq:quenchspectext}
\end{equation}
and the \RAA\ is:
\begin{equation}
\begin{split}
\RAA & = \left(1 + \frac{dS}{d\pTjet}\right) \left(\frac{\mathpTz}{\pTjet}\right)^{2\beta\log{\left(1+ S/\mathpTjet\right)}}
\times \\
& \phantom{{}={}}  \left(\frac{1}{ 1 + S/\mathpTjet}\right)^{n +  
\beta\log{\left(1+ S/\mathpTjet\right)}}.
\end{split}
\label{eq:RAAsingleextpower}
\end{equation}
The logarithmic term in the exponent of the extended power-law
function not only produces a corresponding logarithmic term in the
exponent of $1/(1 + S/\pTjet)$ but it also generates an explicit
dependence on $\pTz/\pTjet$. Thus, even for a constant fractional
shift, the \RAA\ decreases with increasing \pTjet\ for positive
$\beta$: 
\begin{equation}
\RAA = \left(\frac{\mathpTz}{\mathpTjet}\right)^{2\beta\log{\left(1 + s\right)}}
\left(\frac{1}{ 1 + s}\right)^{n - 1 +  
\beta\log{\left(1+ s\right)}}.
\end{equation}
For a spectrum consisting of both quarks and gluons, the \RAA\ is
given by an expression similar to Eq.~\ref{eq:raaqg} containing an
$f_q$- and $(1-f_q)$-weighted combinations of
Eq.~\ref{eq:quenchspectext} with different values for $n$ and
$\beta$ for quarks and gluons. Given the $f_q$ distributions in
Fig.~\ref{fig:fqpt}, a constant fraction shift is unable to reproduce
the measured increase of the jet \RAA\ with increasing \pTjet. In
fact, a calculation of the \RAA\ using the extended power-law form
well reproduces the results of the Monte Carlo evaluation shown in
Fig.~\ref{fig:raacomparepower}. Thus, we conclude that the apparent
success of the constant fractional shift scenario in
Fig.~\ref{fig:raacomparepower} using the analytic analysis is false
and results from neglecting the deviations of the jet spectra
from the pure power-law form. 

\section{Modeling jet \RAA and \Dz: non-constant fractional shift}
\label{sec:RAADZNonConstantfracShift}
The inability of the constant fractional shift assumption to explain
the \pTjet\ dependence of the measured \RAA\ suggests that the shift
has to vary with \pTjet\ more slowly than linearly. In fact, 
theoretical analyses of medium-induced energy loss would not be
compatible with a constant fractional shift scenario. To proceed, we
make the minimal extension of the analysis above and assume that the
shift is proportional to an undetermined power of \pTjet,
\begin{equation}
S = s' \left(\frac{\mathpTjet}{\mathpTz} \right)^\alpha.
\label{eq:nonconstshift}
\end{equation}
Lacking any knowledge of the appropriate scale for the term in the
parenthesis, we use the same reference scale, \pTz, used for
parameterizing the spectra. Then, $s'$, which has dimensions of energy
or transverse momentum, represents the shift in transverse
momentum for jets having $\pTjet = \pTz$. 
The resulting \RAA\ for a single jet spectrum and for combined
quark and gluon spectra can be obtained using the procedures described
above, in particular a combination of Eq.~\ref{eq:quenchspectext}
weighted by the \pTjet-dependent quark and gluon fractions; the
formulas are not shown here for sake of brevity.

\begin{figure*}[t]
\centerline{
\includegraphics[width=0.9\textwidth]{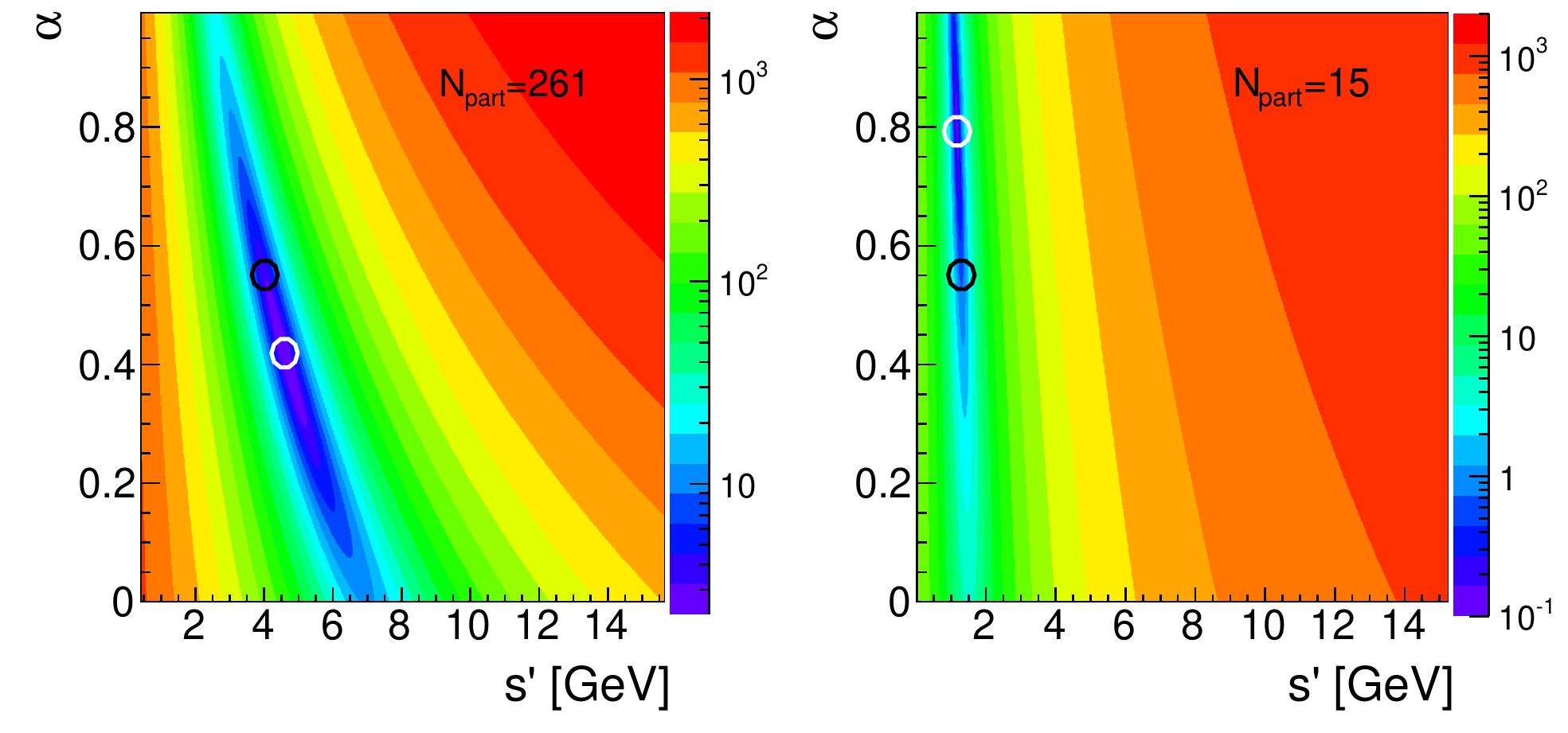}
}\
\hspace{0.6cm}\includegraphics[width=0.875\textwidth]{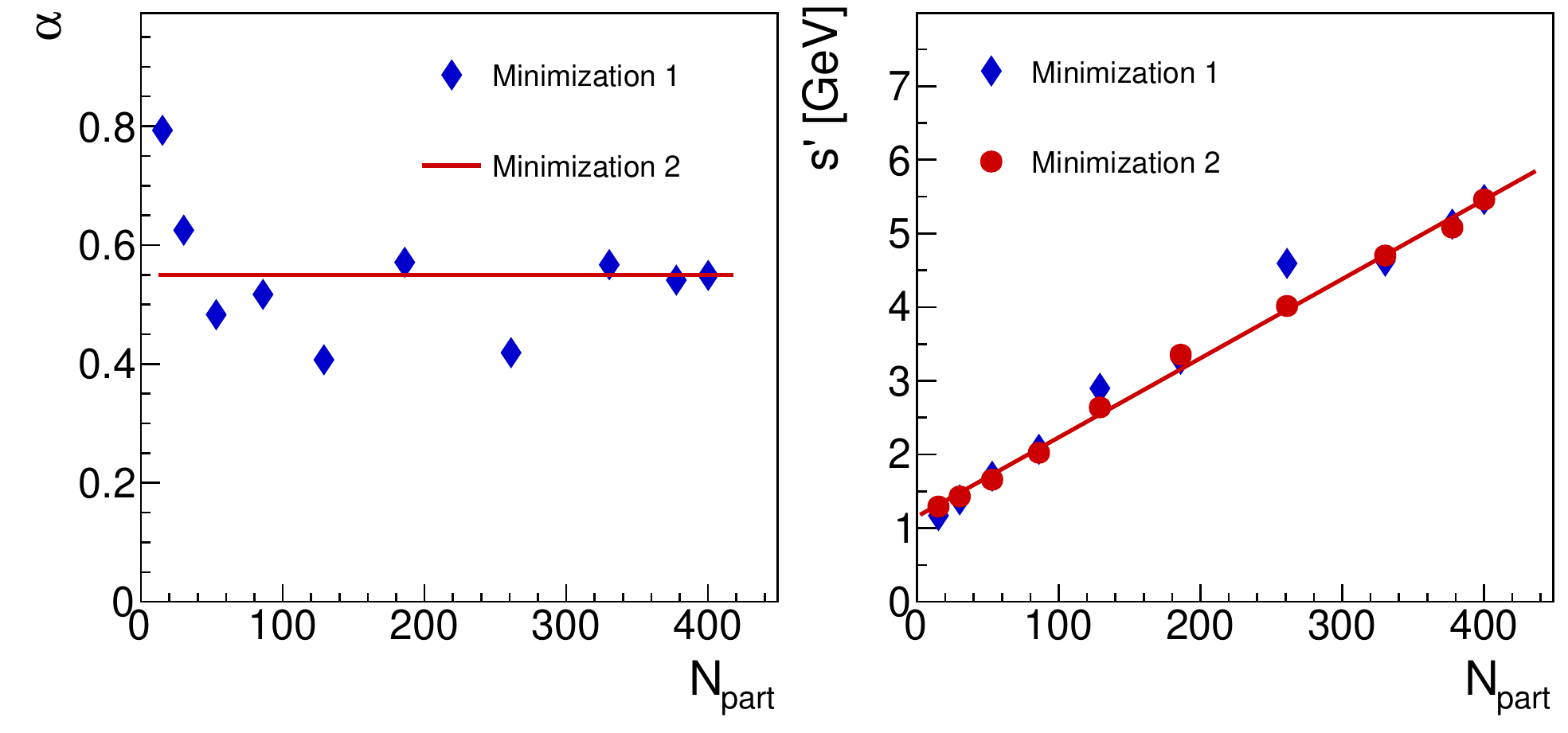}
\caption{
Top:  $\chi^2/{\rm DOF}$ as a function of $\alpha$ and $s'$ for the
10-20\% (left) and 70-80\% (right) centrality bins. The positions of
the minima with $\alpha$ free (fixed) are indicated by the white
(black) circles.
Bottom: Parameters of non-fractional shift (Eq.~\ref{eq:nonconstshift})
model, $\alpha$ (left) and $s'$ (right), as a function of \Npart\
obtained from fits of the resulting calculated  \RAA\ to the ATLAS
data. The blue points indicate results for which both $\alpha$ and
$s'$ are free parameters while the red points indicate the results of
fits with $\alpha$ fixed to 0.55 (see text) shown on the left panel by
the red line. The line on the right
panel shows the result of a linear fit to $s'(\Npart)$.
}
\label{fig:raafits}
\end{figure*}

The shift expression in Eq.~\ref{eq:nonconstshift} and the resulting
\RAA\ were used to perform a fit to the ATLAS data to extract $s'$ and
$\alpha$ in different centrality bins. The fits were performed 
using the statistical uncertainties in the $\chi^2$. The results are
presented in Fig.~\ref{fig:raafits} where the top panels show $\chi^2$
contours in $s'$--$\alpha$ space for the 10-20\% (left) and 70-80\%
(right) centrality bins. As the contours demonstrate, there is a strong
correlation between the parameters which causes the obtained optimal
$\alpha$ and $s'$ values, shown in the lower panels of the figure as a function
of \Npart, to fluctuate. The $\alpha$ values, however, cluster around
an average value of 0.55. To reduce the point-to-point scatter in the
obtained parameters $\alpha$ was fixed to the value 0.55 and the fits
were run again to extract $s'$. The values are shown in the lower right
panel with the red points. The white and black circles shown on the
$\chi^2$ contour plots indicate, respectively the results of the free
fits and the fits with $\alpha = 0.55$. 

The $s'$ values obtained using fixed $\alpha$ show an approximately
linear dependence on \Npart\ and vary from $\sim 1$~GeV in the most
peripheral bin to $\sim 5.5$~GeV in the most central (0-1\%) bin. The
fact that $s'$ extrapolates to a non-zero value for $\Npart
\rightarrow 0$ may indicate that there is an additional contribution
to the measured single-jet suppression present in even peripheral
collisions.  In fact, the free fits suggest a systematic rise in
$\alpha$ for the two most peripheral bins which may arise from the
same underlying physics.

\begin{figure*}[t]
\begin{center}
\includegraphics[width=0.8\textwidth]{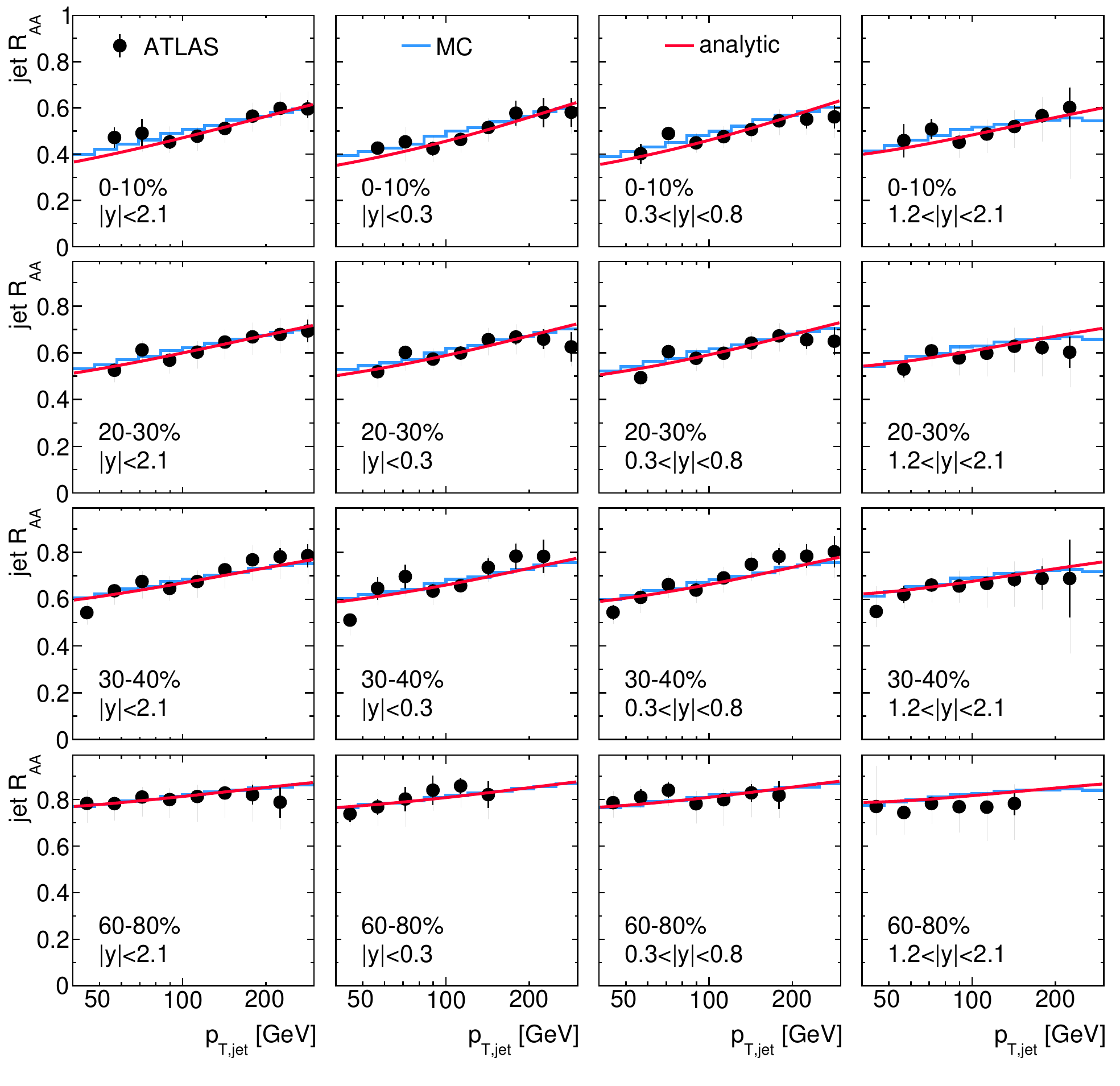}
\end{center}
\caption{
 Nuclear modification factor of jets, \Raa, measured by ATLAS \cite{Aad:2014bxa}
 (black markers) in four different centrality bins (rows) and four
 different rapidity regions (columns) is compared to the analytic
 calculation (red line) and MC calculation (blue histogram) of the
 same quantity in the non-constant fractional energy loss model. The
 analytic calculation uses the extended power law parameterization of
 the jet \pt\ spectra that includes the logarithmic dependence of the
 exponent on jet \pt . 
  }
\label{fig:Fig5}
\end{figure*}

The \RAA\ calculated using the results of the fitting procedure are
shown in Fig.~\ref{fig:Fig5}. Using the extended power-law
parameterization of the quark and gluon spectra, the analytic and
Monte Carlo results are in good agreement. The growth of the
\RAA\ with \pTjet\ results from the fact that the shift increases with
\pTjet\ more slowly than linearly. Thus, the fractional energy loss
decreases with increasing \pTjet. The agreement with the data is
largely by construction since the parameters of the energy loss were
obtained from the above-described fitting procedure. Nonetheless, our
model is capable of describing the available data with a
single, centrality-independent value for $\alpha$ and a proportionality
constant, $s'$ that varies approximately linearly with \Npart. 

\begin{figure}[t]
\begin{center}
\includegraphics[width=0.95\fighalfwidth]{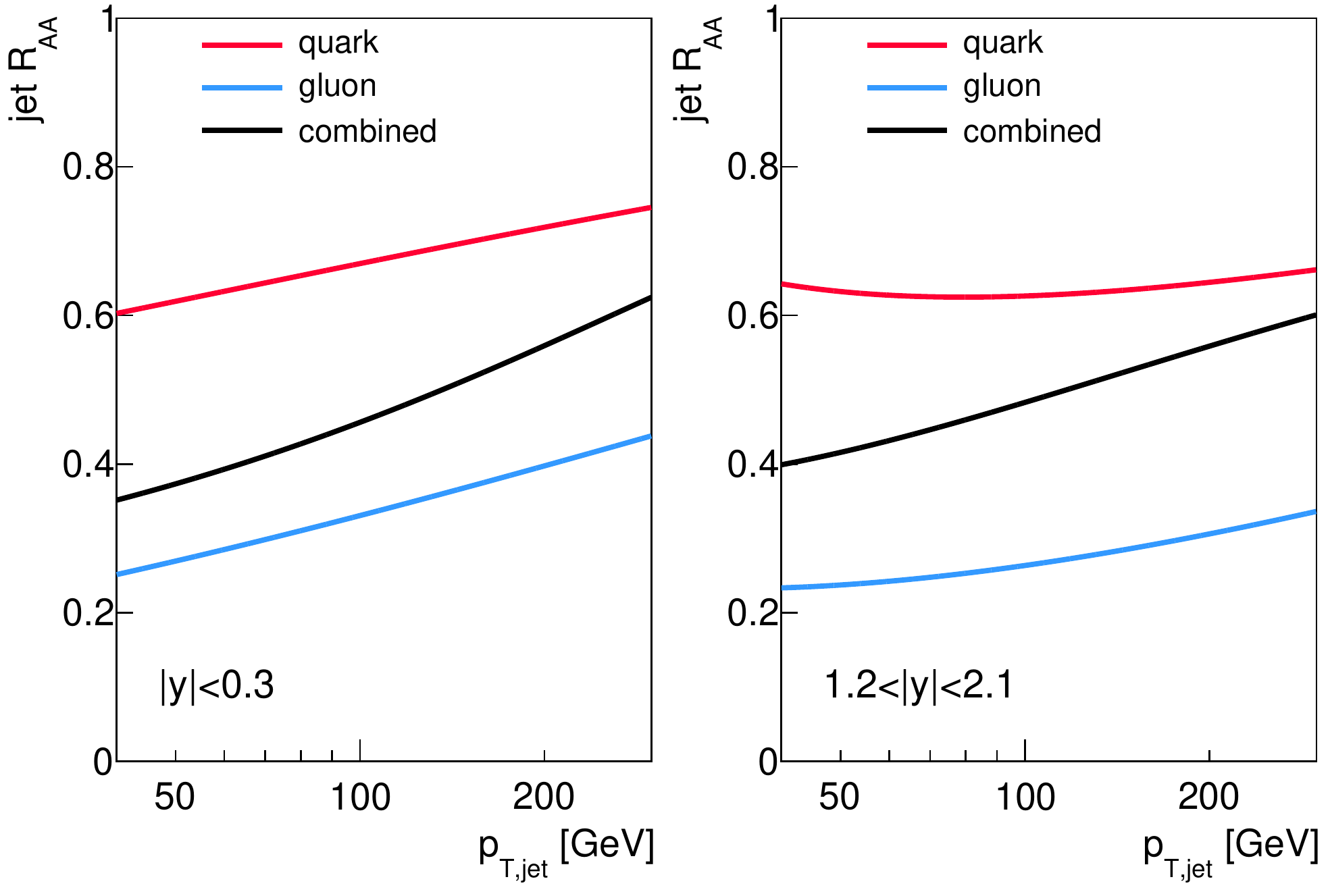}
\end{center}
\caption{
Quark, gluon, and combined \RAA\ vs \pTjet\ for the $|y|<0.3$ (left) and $1.2 < |y| <
2.12$ (right) rapidity bins.  
 }
\label{fig:fig17}
\end{figure}

The
rapidity dependence, or lack thereof, in the \RAA\ arises from a
cancellation between the rapidity dependence of the quark fraction,
which increases with increasing rapidity (see
Fig.~\ref{fig:fqpt}), and the shapes of the quark and gluon
spectra which become steeper with increasing rapidity. 
This cancellation is illustrated in Fig.~\ref{fig:fig17} which shows
the suppression for the quark, gluon, and combined spectra in the 
$|y|<0.3$ (left) and $1.2 < |y| < 2.1$ (right) rapidity bins.  The
difference between the quark and gluon suppression is greater in the 
$1.2 < |y| < 2.1$ than in the $|y| < 0.3$ bin. The \pTjet\ dependence
is also much flatter in the higher rapidity bin. Yet, the combined
suppression taking into account the \pTjet\ dependence of
$f_{\mathrm{q}}$ in Fig.~\ref{fig:fqpt} is nearly the same in the two
rapidity bins. 

\begin{figure*}[t]
\begin{center}
\includegraphics[width=0.9\textwidth]{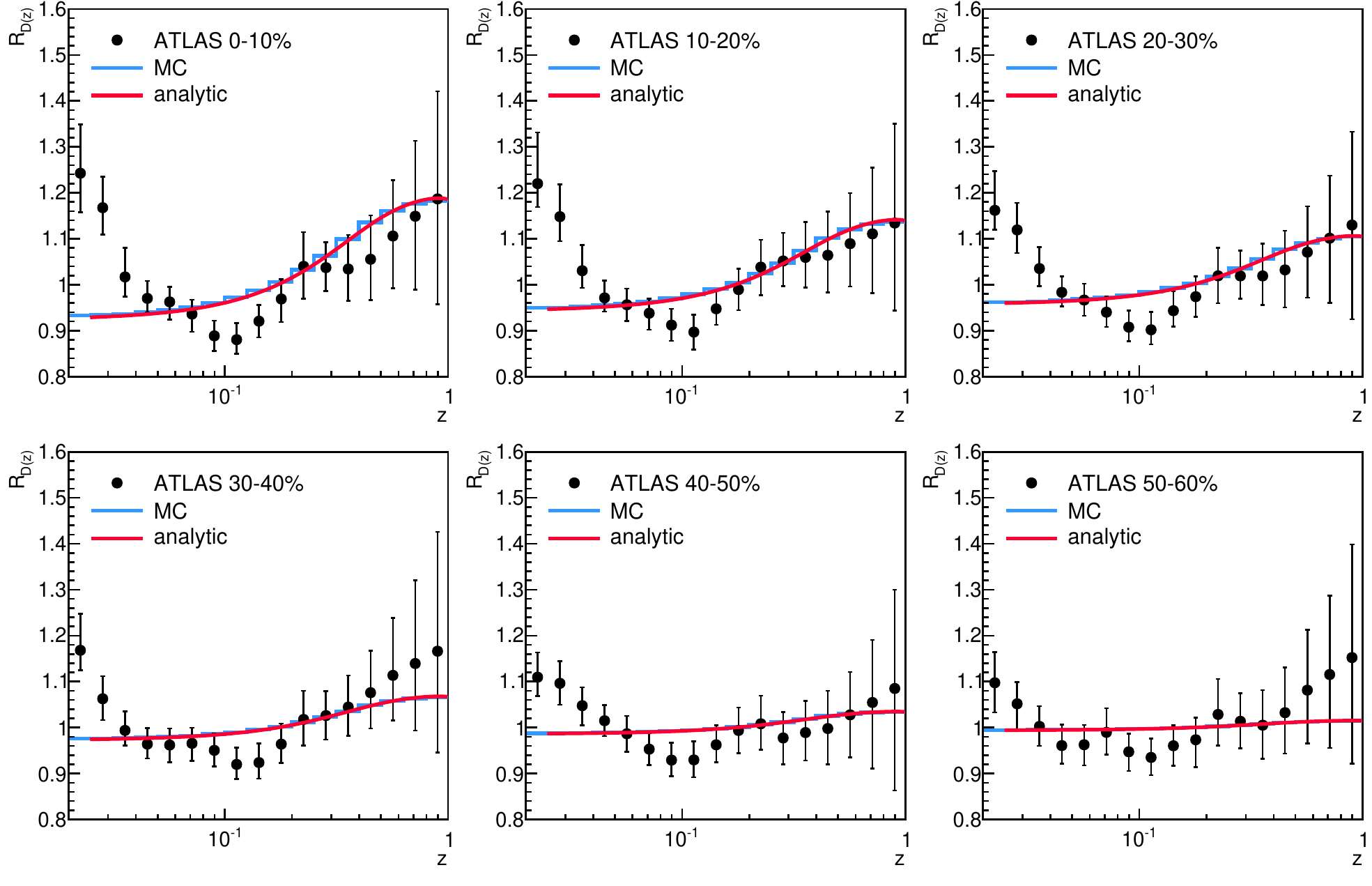}
\caption{
  Ratios of $D(z)$ distributions for six bins in collision centrality to those in peripheral (60-80\%) collisions, $D(z)|_{\mathrm{cent}} /D(z)|_{60-80}$, measured by 
ATLAS for $R=0.4$ jets \cite{Aad:2014wha} (black markers) are compared to the analytic calculation (red line) and MC calculation (blue histogram) of the same quantity in the 
non-constant fractional energy loss model.
  }
\label{fig:Fig9}
\end{center}
\end{figure*}

The \Dz\ distributions calculated using the extended power-law
functions and the shift in Eq.~\ref{eq:nonconstshift} are shown in
Fig.~\ref{fig:Fig9}. As with the \RAA, the agreement between the
analytic calculation and the Monte-Carlo sampled result is much better
using the extended power-law descriptions of the primordial parton
spectra. However, the \Dz\ modifications in the model are largely the
same using the fractional and non-fractional shift parameterizations. 
This lack of sensitivity to $S(\pTjet)$ arises because the
\Dz\ measurements are dominated by contributions from jets with
$\pTjet \sim 100$~GeV and because the \Dz\ modifications in the model
primarily result from the difference between quark and gluon quenching
for jets with similar transverse momenta. Thus, as long as the model
reproduces the \RAA\ near 100~GeV the \Dz\ modifications will be
insensitive to the \pTjet\ dependence of $S$.

\section{Rapidity dependence of the suppression}
\label{sec:Forward}

 The fraction of jets initiated by light quarks evolves as a function
 of the rapidity such that the probability that the jet is
 initiated by a quark is increasing with increasing
 rapidity. The steepness of the jet \pt\ spectrum also evolves
 as a faction of the rapidity such that the \pt\ spectra of
 forward jets are  steeper than the spectra of jets produced in the
 central region. Both of these features are demonstrated in
 Fig.~\ref{fig:fqpt} and in Table~\ref{tab:fitparams} of
 Sec.~\ref{sec:JetSpectra}.  Both features also influence the jet
 \Raa, though they act in opposite directions. Nonetheless, it can
 reasonably be expected that the jet \Raa\ will exhibit a
 different behavior in the forward region compared to the central
 region, or, equivalently, that the \Raa\ will vary with rapidity at
 sufficiently large values. Thus, it is clearly of interest to test
 the model presented in this 
 paper by predicting the jet \Raa\ in the forward region where it has
 not yet been measured. To do that,
 the jet \Raa\ was  calculated using the analytic model in two bins of
 jet rapidity corresponding to those used by ATLAS or CMS
 \cite{Aad:2014vwa,Chatrchyan:2012gwa}, namely  $2.1<|y|<2.8$ and
 $2.8<|y|<3.5$. In the later rapidity region, the jet
 \pt\ spectra decrease approximately by four orders of magnitude in
 the region of jet  \pt\ between 40~GeV and 100~GeV. This steep
 fall-off of the spectra was found to be insufficiently described by
 the modified power-law, Eq.~\ref{eq:extpowerlaw}. To improve the
 parameterization, an additional quadratic term was introduced leading
 to the parameterization, 
  \begin{equation}
  \frac{dn}{d\pTjet} = A \left(\frac{\pTz}{\pTjet}\right)^{n + \beta \log{\left( \mathpTjet/\mathpTz \right)} + \gamma \log^2{\left( \mathpTjet/\mathpTz \right) }},
  \label{eq:extextpowerlaw}
  \end{equation}
  which was found to describe the PYTHIA jet \pt\ spectra at the level
  of accuracy better then 10\%. The resulting parameters and the quark
  fractions for the jet \pt\  spectra selected in the two
  rapidity regions are summarized in
  Table~\ref{tab:fitparamsforward}. 

The resulting analytic \Raa\ was calculated using an extension of
Eq.~\ref{eq:RAAsingleextpower} to account for the quadratic term,
and using the results from 
  Sec.~\ref{sec:RAADZConstantfracShift}, namely a shift of the form of
  Eq.~\ref{eq:nonconstshift} with $\alpha = 0.55$ and $s'(\Npart)$ as
  shown in Fig.~\ref{fig:raafits}. 
\begin{table}
\begin{center}
\begin{tabular}{|l||c|c|c|c|} \hline
Parameter       & $2.1<|y|<2.8$ & $2.8<|y| < 3.5$ \\ \hline  \hline
 $n_{q}$         & 5.5 & 6.7   \\ \hline 
 $n_{g} $        & 6.3 & 7.4   \\ \hline 
 $\beta_{q}$    & 0.34 & -0.46   \\ \hline 
 $\beta_{g}$    & 0.52 & -1.19   \\ \hline 
 $\gamma_{q}$   & 1.2 & 2.6   \\ \hline 
 $\gamma_{g}$   & 1.5 & 2.4   \\ \hline 
 $ f_{q} $       & 0.60 & 0.76   \\ \hline 
\end{tabular}
\end{center}
\caption{Parameters obtained from fits of the \PYTHIA\ forward jet spectra 
to
the extended power-law (Eq.~\ref{eq:extextpowerlaw}) forms.
}
\label{tab:fitparamsforward}
\end{table}

\begin{figure}[htb]
\begin{center}
\includegraphics[width=0.9\fighalfwidth]{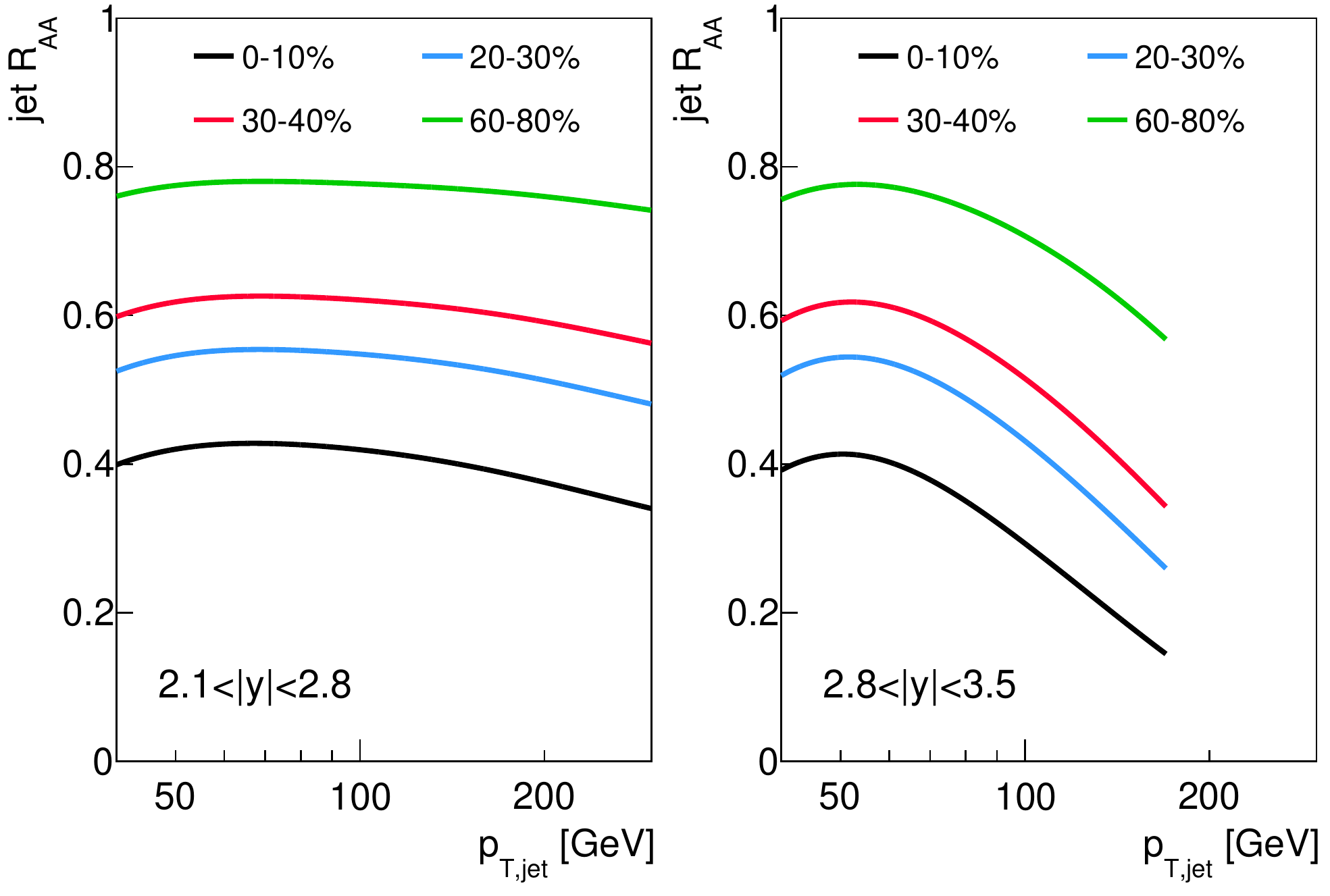}
\caption{
  Predicted jet \Raa\ as a function of \pTjet\ in forward rapidity
  intervals $2.1<|y|<2.8$ (left) and $2.8<|y|<3.5$ (right) shown
  for four centrality bins, 0-10\% (black), 20-30\% (blue), 30-40\%
  (red), and 60-80\% (green). See text for details.
  }
\label{fig:Fig8}
\end{center}
\end{figure}

  The predicted forward \Raa\ is shown as a function of \pTjet\ in
  Fig.~\ref{fig:Fig8}. A clear change 
  in the trend of the \Raa\ evolution with jet \pt\ can be
  seen. In contrast to the slow increase seen for the jet \Raa\ in the
  rapidity regions within $|y| < 2.1$, the jet \Raa\ in the
  forward regions first increases, reaches a maximum and then 
  decreases with increasing \pTjet.  The decrease is more pronounced for more
  forward region where the jet \Raa\ in 0-10\% central collisions
  reaches the maximum of approximately 0.4 at around 50~GeV, and then it decreases
  reaching a value of approximately 0.15 at 170~GeV. These trends are
  present across different centralities. Such pronounced change in the
  behavior of the forward jet \Raa\ represents a distinct feature
  that can be tested by future measurements at the LHC.

\begin{figure*}[t]
\begin{center}
\includegraphics[width=0.9\textwidth]{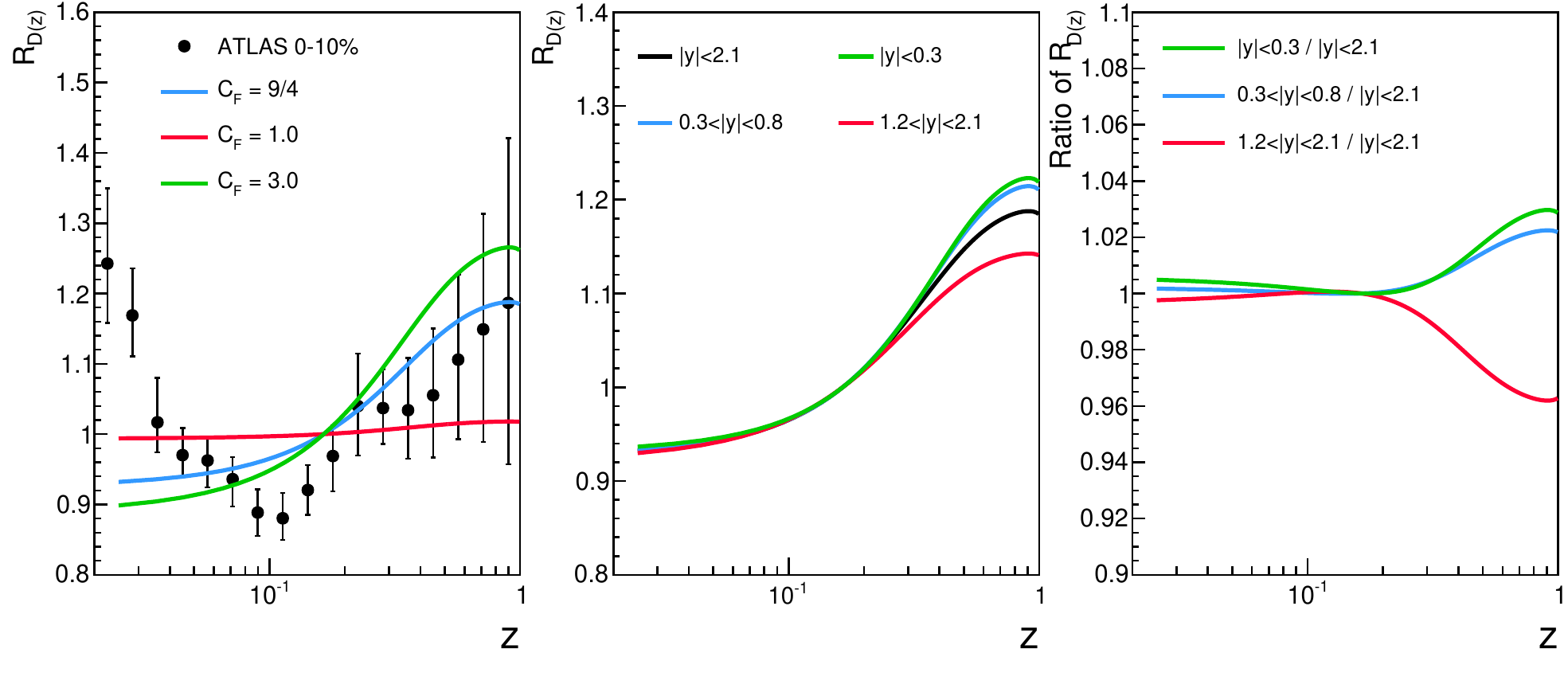}
\caption{
  Left panel: ATLAS data on the ratio of $D(z)$ distributions,
  $R_{D(z)}$, in 0-10\% to 60-80\% centrality bin (black markers)
  compared to the analytic calculation of the non-constant fractional
  energy loss for three different values of color factors: the default
  value of color factor $C_F=9/4$ (blue), $C_F=1.0$ (red), and
  $C_F=3.0$ (green). Middle panel: Prediction of the ratio of $D(z)$
  distributions, $R_{D(z)}$, in 0-10\% to 60-80\% centrality bin in
  different jet rapidities. The $R_{D(z)}$ in the rapidity
  region of $|y|<2.1$ (black) is compared to the $R_{D(z)}$ in the
  regions: $|y|<0.3$ (green), $0.3<|y|<0.8$ (blue), and
  $1.2<|y|<2.1$ (red). Right panel: The ratio of predicted $R_{D(z)}$ in
  different rapidity regions to the $R_{D(z)}$ in the region of
  $|y|<2.1$. 
  }
\label{fig:Fig10}
\end{center}
\end{figure*}

  The dependence of the quark fraction on the jet rapidity has
  to influence also the trends measured in the centrality dependent
  ratios of fragmentation functions, \Rdz, presented in
  Sec.~\ref{sec:RAADZNonConstantfracShift}. As demonstrated in
  Sec.~\ref{sec:RAADZConstantfracShift}, this ratio exhibits only a
  weak dependence on the shape of the underlying jet \pt\ spectra and
  thus, it is a very useful observable that may help isolate 
  the effects of different quenching of quark-initiated and
  gluon-initiated jets. The sensitivity of the \Dz\ modification to
  differences in quark and gluon quenching are illustrated in the left
  panel of Fig.~\ref{fig:Fig10} which shows \Rdz\ in 0-10\%
  central collisions evaluated for three different choices of an
  effective color factor, $C_F = 1.0, 9/4$, and 3.0. A clear dependence of the
  \Rdz\ on the color factor can be seen. For equal suppression of the
  quark-initiated and gluon-initiated jets, the \Rdz\ exhibits only
  negligible difference from unity reflecting minor difference in the
  quark fraction at unquenched and quenched jet transverse
  momenta. For color factor larger than the default value of $9/4$,
  the \Rdz\ exhibits larger increase compared to the \Rdz\ evaluated
  with the default value of $C_F$.

  The middle panel of Fig.~\ref{fig:Fig10} shows the \Rdz\ evaluated
  in 0-10\% central collisions for the rapidity region $|y| <
  2.1$ and the prediction for the rapidity dependence of the
  \Rdz\ for three rapidity regions chosen to match the
  rapidity regions used in the measurement of the jet \Raa\ by
  ATLAS \cite{Aad:2014bxa}, namely $|y|<0.3$, $0.3<|y|<0.8$, and
  $1.2<|y|<2.1$. The differences in the \Rdz\ are better quantified
  in terms of the ratio of \Rdz\ evaluated in a given rapidity
  region to that evaluated in the region of $|y| < 2.1$ as shown in
  the right panel of Fig.~\ref{fig:Fig10}. The ratio is predicted to
  reach larger values in the more central rapidity region and
  smaller values in more forward region compared to the inclusive
  rapidity interval. Both of these effects are in the maximum at
  the level of 6-7\% of the inclusive \Rdz .  This predicted behavior
  represents another possibility to test the model and, more
  generally, to probe the differences between the energy loss of quark
  and gluon jets.

\section{Modeling $D(\pt)$ and charged particle \Raa}
\label{sec:Particles}

\renewcommand{\Dpt}{\mbox{$D(p_{\mathrm{T}}^{\mathrm{ch}})$}} 
\newcommand{\mathDpt}{D(p_{\mathrm{T}}^{\mathrm{ch}})} 
\newcommand{\Raach}{\mbox{$R_{\rm AA}^{\rm ch}$}} 
\newcommand{\RDpt}{\mbox{$R_{D(p_{\mathrm{T}}^{\mathrm{ch}})}$}} 

We have shown that our simple model for the medium modifications of 
single jets can reproduce the measured nuclear modification factor and 
its rapidity and transverse momentum dependence.  The model can also 
reproduce some of the qualitative features seen in measured inclusive 
jet fragmentation functions, namely the suppression at intermediate $z$ 
and an enhancement at large $z$. It was shown that these features in the 
measured fragmentation functions arise from the change in the quark (or 
gluon) fraction due to the medium-induced energy loss. Since the model 
can explain both the jet spectra and the fragmentation functions, it 
should be also able to reproduce the charged particle transverse 
momentum distribution, \Dpt, of charged particles produced within jets, 
and the nuclear modification factor of charged particles, \Raach , 
measured at high transverse momenta of charged particles, \ptch. The 
\Dpt\ distributions were measured by ATLAS \cite{Aad:2014wha} and CMS 
\cite{Chatrchyan:2014ava} for jets with $\pTjet > 100$~GeV. The charged 
particle \Raach\ at high \ptch\ was measured by CMS \cite{CMS:2012aa}.  
Compared to the inclusive jet fragmentation functions which are largely 
independent of the jet transverse momentum for a given flavor of the 
initial parton, these observables couple together the change in the jet 
fragmentation and the change in the underlying jet spectra. Thus, these 
observables provide another important input to test the model.

\begin{figure*}
\begin{center}
\includegraphics[width=0.8\textwidth]{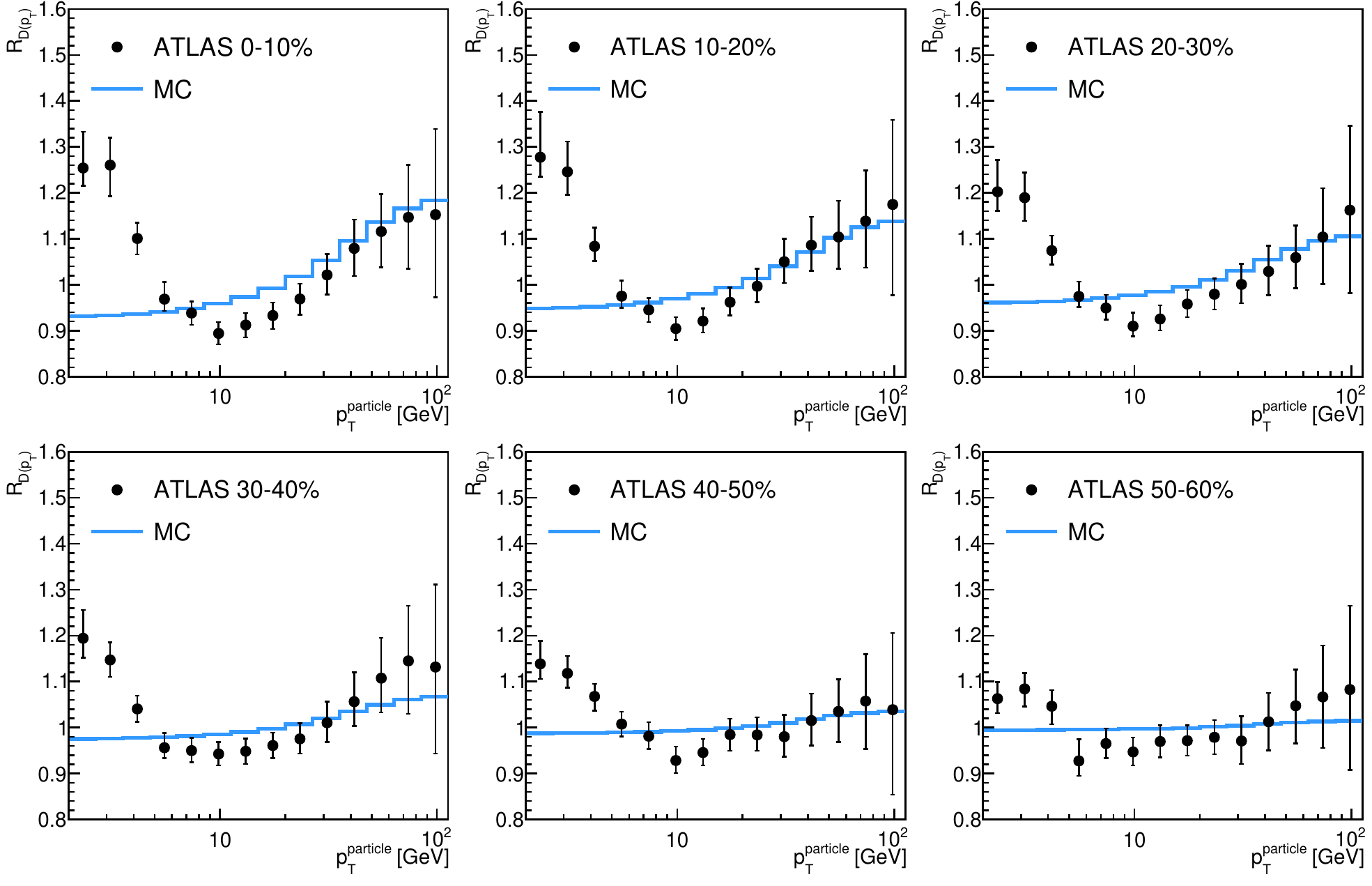}
\caption{
  Ratios of $\Dpt$ distributions for six bins in collision centrality 
to those in peripheral (60-80\%) collisions, $\mathDpt|_{\mathrm{cent}} 
/\mathDpt|_{60-80}$, measured by ATLAS for $R=0.4$ jets \cite{Aad:2014wha} (black 
points) are compared to the MC calculation (blue histogram) of the same 
quantity in the non-constant fractional energy loss model.
  }
\label{fig:Fig11}
\end{center}
\end{figure*}

 The PYTHIA8 simulated events were used to simulate the modifications 
of the \Dpt\ distributions. The simulation was done in two steps. In the 
first step, the \Dpt\ distributions were booked in 1~GeV bins of the 
\ptjet\ to two look-up tables, for quark initiated and gluon 
initiated jets separately. In the second step, the jet suppression was applied at 
the single jet level in the same way as for the case of simulating the 
inclusive jet \Raa . For each suppressed jet of a given flavor, the 
\Dpt\ distribution corresponding to the quenched jet \pt\ was read off 
from the look-up table and added to the histogram which formed the 
resulting \Dpt\ distribution after being properly normalized by a total number of 
quenched jets with $\pTjet > 100$~GeV. The central to peripheral ratio of 
\Dpt\ distributions, \RDpt , was evaluated. The result is shown along 
with the data by ATLAS in Fig.~\ref{fig:Fig11}. The figure shows that our model 
can reproduce the qualitative features observed in the data,
namely the suppression of yields at intermediate \ptch\ and an 
enhancement at high \ptch. This is not surprising given the success of
the model in describing the \Dz\ modifications, but it represents an
important consistency check.

\begin{figure*}
\begin{center}
\includegraphics[width=0.8\textwidth]{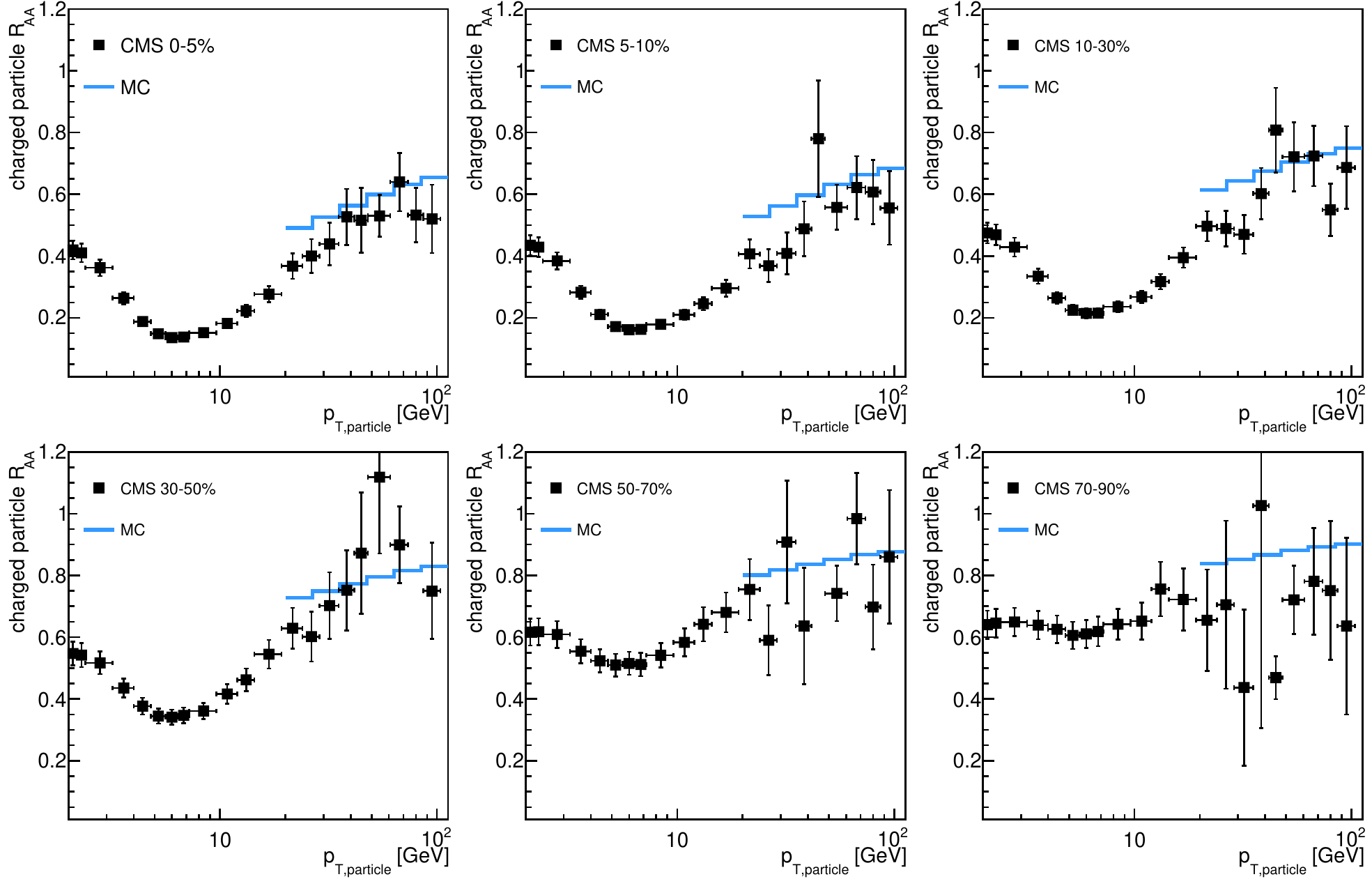}
\caption{
  Nuclear modification factor of charged particles measured by CMS 
\cite{CMS:2012aa} in six centrality bins (black points) is compared to the MC 
calculation of the same quantity in the non-constant fractional energy loss 
model (blue histogram).
  }
\label{fig:Fig12}
\end{center}
\end{figure*}

To evaluate the nuclear modification factor of charged particles, 
\Raach , one can use the same procedure as for evaluating the \RDpt\ 
with only two differences which is evaluating the \Dpt\ distributions 
for jets with no threshold on jet \pt\ and avoiding a normalization of 
the final \Dpt\ distributions by the total number of jets.
  This approach is based on the fact that each charged particle with a 
given \ptch\ must come from the jet with $\pTjet \geq \ptch$. This allows 
to construct the \Raach\ for $\ptch > 20$~GeV which is the kinematic 
region where we have a good confidence in modeling the inclusive jet 
suppression. The resulting \Raach\ is shown along with the \Raach\ 
measured by CMS in Fig.~\ref{fig:Fig12}. 

The figure shows that the model 
can reproduce the qualitative features seen in the data at 
high-\ptch , namely the increase of the \Raach\ with increasing \ptch\ 
and its centrality dependence. However, the slope of the charged
particle \RAA\ in the model  differs from that in the data at the
lowest \pTch\ values included in this analysis and the model
systematically slightly over-predicts the \Raach\ in the data.
These disagreements may be due to insufficient precision in the 
modeling of \Dpt\ distributions by PYTHIA8, something that
can and will be tested in future analyses. There may be consequences
from the disagreements between the data and the model in the \Rdz\ 
(\RDpt) including the low-$z$ (low-\ptch) excess and fact that the
data is  systematically lower than the model in the range $0.1 < z <
0.2$ ($10 < \ptch < 20$~GeV). 
  The former is discussed in the next section, the latter may be
insufficient to explain the observed difference.
  Generally, the disagreement may be due to jet quenching physics not included in the
model. In particular, the possibility that the jet quenching produces
changes in the fragmentation functions of lower-\pTjet\ jets different from
that observed in the $\pTjet > 100$~GeV jets cannot be excluded.

\section{\Dz\ low-$z$ excess}
As described above, our model for the modification of the
fragmentation function in \PbPb\ collisions is based on the assumption
that energy lost during the evolution of a parton shower in the medium
does not appear as part of the measured jet and that the final fragmentation
products have the same \Dz\ distribution as if the reduced-energy jet
fragmented in vacuum. While the validity of these assumptions may be
debated (next section), the soft excess in the
\PbPb\ \Dz\ distributions, which cannot be explained by the different
quark and gluon quenching, presents a manifest violation of the
assumptions of the model. Supposing that the enhancement at low-$z$
reflects either radiation within the jet, recoil partons, or
collective response of the medium, from the point of view of our
model, the extra low-$z$ hadrons provide an upward shift of the jet
energy. In the ATLAS \PbPb\ fragmentation function measurement, the
excess fragments in the range $0.02 < z < 0.04$ were found to
contribute $\sim 2\%$ of the jets transverse momentum for the 0-10\%
centrality bin. Since that estimate could not account for
contributions from hadrons below the minimum transverse momentum of
the charged particle measurement, it is clearly an under-estimate of
the contribution of ``excess'' low-$z$ partons to the jets energy.
However, we take this number to be an order-of-magnitude estimate of
the fractional contribution of the low-$z$ excess to the energy of the
typical jet which we will refer to as \fsoft. 

\begin{figure*}[t]
\centerline{
\includegraphics[width=0.9\textwidth]{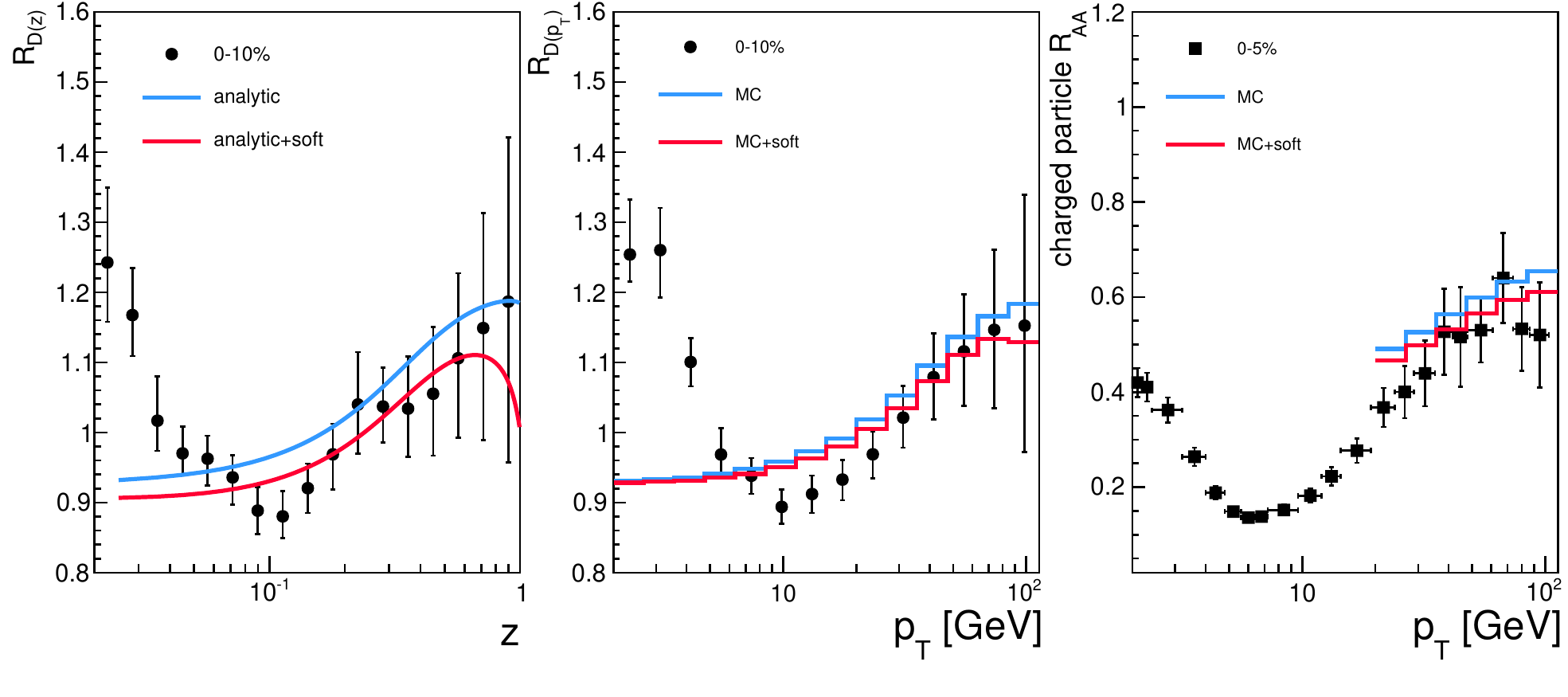}
}
\caption{
Demonstration of the effects of accounting for the contribution of
low-$z$ particles in jet energies (see text). Left: calculated
\Rdz\ with (red) and without (blue) the low-$z$ excess; middle: same
for \Rdpt; right: \Raach\ with and without the soft-$z$ excess.
}
\label{fig:softq}
\end{figure*}
Since the energy in the low-$z$ fragments likely is proportional to
the jets energy, we assume that 
\begin{equation}
\fsoftinc = \fqint \fsoftq + ( 1 -\fqint ) \fsoftg, 
\label{eq:soft}
\end{equation}
with $\fsoftg = \cf \fsoftq$ and where \fsoftq, \fsoftg, and
\fsoftinc\ represent the average soft 
excess in quark, gluon, and all jets, respectively. The contribution
of the excess soft hadrons to the jet energy is, in 
principle, already effectively accounted for in our analysis of the
jet suppression where it will reduce $S$. However, it could affect
our description of the modified fragmentation functions by increasing
the jet energy that appears in the denominator of the $z$ definition
in the data by a fraction $1 + \fsoft$, and, thus, reducing the $z$
values by a factor $1/\left( 1 + \fsoft\right)$. Thus, the modification of the
fragmentation function in the data at $z$
should correspond to the modification in our model at a $z$ value of
$z\left( 1 + \fsoftq\right)$. To reproduce that effect, we have
corrected Eq.~\ref{eq:dzmodpower} as follows,
\begin{equation}
\begin{split}
D^{\mathrm{meas}}(z) & = f_q^{\mathrm{int}} D_q\left(z \left[1+\fsoftq\right] \right) + \\
& \phantom{{}={}} \left(1 - f_q^{\mathrm{int}}) D_g(z \left[1+\fsoftg\right] \right),
\end{split}
\end{equation}
and recalculated the \Dz\ modifications taking $\fsoftinc = 0.02$. The
results are shown in the left panel of
Fig.~\ref{fig:softq}. 
 Accounting for the shift in the jet energy shifts \Rdz\ down and leads 
to a better agreement between the data and the model. More than one 
sigma disagreement my now be seen only for the data point near $z=0.1$.

The depletion seen in the modeled $R_{D(z)}$ at high-$z$ reflects the 
rapid decrease in the parameterized \Dz\ for $z\gtrsim 1$. The result
in Fig.~\ref{fig:softq} is continuous across $z = 1$ for reasons given
in Sec.~\ref{sec:JetSpectra}, but the \Dz\ still falls rapidly near $z =
1$. The data do not yet have the precision to resolve a change in
behavior near $z = 1$ like that shown in the figure. Future
measurements with improved precision could test for such an
effect. Observing the depletion would provide strong empirical support
to the picture that (most) jets fragment as in the vacuum but with additional
energy from low-$z$ particles whose origin is not yet understood. 

  The contribution of the excess soft hadrons to the jet momentum also 
influences the modeled \RDpt\ and \Raach . The impact of this 
contribution can be avoided by reducing the quenched jet momentum used 
to look-up for the \Dpt\ distributions corresponding to quenched jets 
by $\fsoft$. The impact of this on modeled \RDpt\ and \Raach\ for 
$\fsoft = 0.02$ is shown in the middle and right panel of 
Fig.~\ref{fig:softq}, respectively. While the impact of soft excess 
hadrons on the \Dpt\ distribution is rather small the impact on 
\Raach\ is more significant leading to a better agreement of modeled 
\Raach\ with the data.

\section{Discussion}
This paper has presented an analysis of implications of recent data on
single jet suppression and inclusive jet fragmentation in
\PbPb\ collisions. The analysis was based on a simple model for the
quenching of a parton shower in the medium, namely that the parton
shower loses energy to the medium in a manner such that the lost energy
does not appear within the jet ``cone''. The fragmentation of the
resulting jet was assumed to be the same as the fragmentation of a jet
in vacuum. These assumptions may seem unreasonably simplistic given
the current understanding of the evolution of high-energy
parton showers in vacuum where perturbative calculations can describe
many features of the resulting distributions of fragments.
However, studies of the impact of the medium on the
color-coherence of parton showers \cite{CasalderreySolana:2012ef} suggest that
the medium is unable to resolve the internal structure of many jets. 
According to Ref.~\cite{Blaizot:2013hx} those jets interact with the
medium as if they consist of a single color charge, and the
reduced-energy jet fragments as if it were in vacuum. An extensive
analysis of the angular and longitudinal momentum distribution of the
radiated energy
\cite{Blaizot:2012fh,Blaizot:2013hx,Blaizot:2013vha,Blaizot:2014rla,Blaizot:2015jea,Blaizot:2015lma} 
indicates that the medium-induced radiation flow to large angles
consistent with CMS measurements \cite{Casalderrey-Solana:2014bpa}.
The combination of these two idea/results suggests a picture for
jet-medium interactions that is qualitatively similar to that used in
this analysis. 

This work started from a simple hypothesis that the \pTjet\ dependence
of the measured jet \RAA\ and the modifications to the jet
fragmentation functions could both be explained by the different
quenching of quarks and gluons and the \pTjet\ dependence of the
primordial quark fraction. Indeed, the first result using analytic
expressions, power-law spectra and a constant fractional shift
provided remarkably good agreement with both the \RAA\ and the
\Dz\ data. However, that success was short-lived as it was found to
disagree with a Monte-Carlo implementation that used the
\PYTHIA\ quark and gluon spectra directly. When deviations of the
spectrum from the pure power-law form were accounted for, the
discrepancy between analytic and Monte-Carlo results was resolved, but
the constant-fractional shift could no longer describe the data. The
sensitivity of the jet \RAA\ to the shape of the jet spectrum is well
known, but this analysis provides clear demonstration of the
sensitivity of interpretations of the \RAA\ measurements to the
accuracy in the description of the primordial jet spectra.

The analysis presented in this paper appears to rule out the
possibility of a constant fractional shift parameterization of the
effects of quenching on the jet spectrum. While a constant fractional
shift is incompatible with most energy loss calculations,  the weak
\pTjet\ dependence of the measured jet 
suppression has often been (informally) interpreted as indicating
fractional energy loss. However, as shown in
Fig.~\ref{fig:raacomparepower}, even when accounting for the fact that
$f_{\mathrm{q}}$ increases with \pTjet, a constant fractional energy
loss produces an \RAA\ that decreases with \pTjet\ due to the fact the
the primordial jet spectrum steepens relative to a pure power law with
increasing \pTjet. The \pTjet-dependent shift extracted from the data,
in fact, varies like $~\sqrt{\mathpTjet}$ as predicted in
Ref.~\cite{Baier:2001yt}. That reference makes clear that the shift
should not necessarily be interpreted as the average energy loss of
the jets. Indeed, the shift approximately parameterizes the
convolution of the energy loss distribution with the primordial jet
spectrum. Thus, direct interpretation of the $s'$ values extracted in
this analysis is not immediately possible. Nonetheless, with the
simple centrality dependence of $s'$ observed in Fig.~\ref{fig:Fig5},
the full \pTjet\ and centrality dependence of the jet suppression can
be accounted for by three parameters, $\alpha$ and the effective
slope and intercept of the \Npart\ dependence of $s'$.

The apparent accidental cancellation of the variation of spectrum
shapes and quark fraction observed here and in
Ref.~\cite{Renk:2014gwa} that leads to the measured constancy of
\RAA\ with rapidity \cite{Aad:2014bxa} is unfortunate as it
reduces the sensitivity of the measurement to features of the energy
loss that could help to test or constrain calculations of medium-induced
energy loss. However, the extensions of our analysis to larger
rapidity indicate that at larger rapidities, the increased curvature
of the primordial jet spectra will provide a different and stronger
variation of \RAA\ with \pTjet. That is particularly true in the $2.8
< |y| < 3.5$ bin where measurements may be difficult in the most central
collisions but could be performed in more peripheral centrality bins
where the effect should still be measurable. 

The analysis in this paper suggests that much or all of the observed
modifications to the jet fragmentation functions in \PbPb\ collisions
can arise from the different quenching of quark and gluon jets -- {\it
  except for the enhancement at low $z$}. This observation could have
important implications for the theoretical understanding of the
quenching physics. For example, in the context of the picture
presented in Ref.~\cite{CasalderreySolana:2012ef} where quenching of
the jets is influenced by the ability of the medium to resolve the
internal structure of the jet, jets that are not resolved by the
medium fragment according to their vacuum fragmentation
functions. But, the measured \Dz\ distributions will still differ from
those in \pp\ or peripheral \PbPb\ collisions due to the different
energy loss of the quarks and gluons. On the other hand, if the
fragmentation functions of the quark and gluon jets are separately
modified, then we would have to conclude that the medium is resolving
the internal structure of the jet. The simplicity of the analysis used
in this paper is not adequate to draw a firm conclusion that the
\Dz\ modifications at intermediate and large $z$ can be completely
attributed to flavor-dependent quenching. However, the analysis shows
the importance of explicitly addressing the effects of different quark
and gluon quenching in future analyses of the fragmentation functions.
Measurements of jet fragmentation in $\gamma$-jet events, where the
jet spectrum has a larger quark fraction would be valuable in
addressing this issue. 

The analysis in this paper suggests that the magnitude and transverse
momentum dependence of charged particle \Raa\ for particles with
$\pt>20$~GeV can largely result from the different quenching of quark
and gluon jets as well.
 The disagreement between the measured charged particle \Raa\ and the
\Raa\ in the model, namely larger suppression and steeper
\Raa\ as a function of \pt\ seen in the data, remains to be understood.
 It may be arising from missing physics in the model, difference
between the measured charged particle spectra and their PYTHIA8
simulation, or from larger contribution of soft particles to quenched
jets at lower \pt . Irrespective of the source of this disagreement,
the result clearly points to the importance of understanding the
measurements involving single particles in the context of fully
reconstructed jets.

As described above, the enhanced production of hadrons at low $z$
observed in data  \cite{Chatrchyan:2014ava,Aad:2014wha} cannot be
explained by the different energy loss of quarks and gluons. It's also
interesting that it could not be explained by a strong-coupling
calculation of energy loss \cite{Casalderrey-Solana:2014wca} but did
arise in a collisional energy-loss scenario also tested in that same
paper, suggesting that the excess could arise from recoiling
constituents of the medium \cite{Zapp:2009ud}. If this explanation is
correct, then the low-$z$ excess is directly probing (part of) the
medium-response to the passage of jets. An alternative explanation was
provided in Ref.~\cite{Blaizot:2015lma} that showed a soft-$z$ excess
can arise when the medium resolves the constituents of the jet core. 
More speculatively, it might be that the excess arises from the
collective response of the medium such as a diffusion wake
\cite{Gubser:2007ga}. Regardless of the explanation, the energy
contributed to the low-$z$ particles acts in the context of our model
as an extra contribution of the energy of the jet. The effect of this
contribution produces modest but noticeable effects on the
fragmentation functions. We note that the persistence of the low-$z$
excess in non-central collisions indicates that it does not result
from systematics in the measurement. Given the potential importance of
the low-$z$ excess, more detailed measurements in non-central events
where the effects of the underlying event are smaller may be
warranted. 

\section{Conclusions}

This paper has presented an analysis of single jet measurements from
\PbPb\ collisions at the LHC using a simple, phenomenological
model. The analysis used quark and gluon spectra obtained from
PYTHIA8 and applied the ``shift'' formulation described by
Baier~{\it et al.} to describe the single-jet suppression. The
modifications to the jet fragmentation functions were assumed to
result from the different quenching of quarks and gluons assuming that
the quenched jets fragment as they do in vacuum. 

Our analysis showed that the transverse momentum dependence of the
quark fraction plays a role in the evolution of the jet \RAA\ with
\pTjet\ as was also observed in Ref.~\cite{Renk:2014gwa}. However, the
curvature of the primordial jet spectrum relative to a pure power-law
also substantially effects the \RAA. The data were found to be
incompatible with a constant fractional shift, $S = s\pTjet$. Fits of
the data using a shift that varied with \pTjet\ as
$\left({\mathpTjet}\right)^{\alpha}$ yielded a value $\alpha \sim 0.55$ compatible
with the prediction in Ref.~\cite{Baier:2001yt}. The measured
modifications of the inclusive jet fragmentation functions can be
largely explained as resulting from the different quenching of quarks
and gluons. More specifically, the suppression of the fragmentation
function at intermediate $z$ and the enhancement in the fragmentation
function at large $z$ reflect the different shapes of the quark and
gluon fragmentation functions and the increase in the quark fraction
due to the greater energy loss of gluon jets. However, our model is
not able to account for the low-$z$ enhancement in the
\PbPb\ fragmentation functions. If our analysis is correct, the
fragmentation function modifications provide a direct test of the
color-charge dependence of jet quenching. The model can explain most
of the observed suppression of the charged particle yield at high
\ptch\ though it slightly under-predicts the amount of suppression and
has a weaker \ptch\ dependence at the lowest \ptch\ values included
in the analysis.

Because the quark fraction varies as a function of
rapidity, both the \RAA\ and the \Dz\ modifications should evolve as a
function of rapidity, though the rapidity-dependent evolution of the
quark and gluon spectra appears to cancel out the effects of the
changing quark fraction over the rapidity range measured in the ATLAS
data. Predictions are made for \RAA\ and \Rdz\ at large rapidity that
would allow the conclusions of this analysis to be tested.

{\bf Acknowledgment:} The work of MS was supported by Charles University 
in Prague, projects PRVOUK P45 and UNCE 204020/2012. BAC's research
was supported by the US Department of Energy Office of Science, Office
of Nuclear Physics under Award No. DE-FG02-86ER40281. We would like to
thank Aaron Angerami for many useful discussions regarding the issues
raised in thie paper.

\bibliography{bibliography}
\bibliographystyle{elsarticle-num}
\end{document}